\documentclass[12pt,a4paper,leqno]{article}%
\usepackage{amsfonts}
\usepackage{amssymb}
\usepackage{graphicx}
\usepackage{setspace}
\usepackage{algorithm,algpseudocode}
\usepackage[round]{natbib}
\usepackage{amsmath}%
\usepackage{amsthm}%
\usepackage{palatino}
\usepackage{paralist}
\usepackage{subfigure}
\usepackage{multirow}
\usepackage{booktabs}
\usepackage{dsfont}
\usepackage{indentfirst}
\usepackage{enumerate}
\usepackage{longtable}
\usepackage{caption}
\usepackage{bbm}
\usepackage{lscape}
\usepackage{float}
\usepackage{pbox}
\usepackage{tikz}
\usepackage{rotating}
\usepackage{xcolor,colortbl}
\usepackage[title]{appendix}
\usepackage[hyperindex,breaklinks]{hyperref}
\hypersetup{colorlinks=true,       
   linkcolor=red,       
    citecolor=blue,        
    filecolor=magenta,      
    urlcolor=cyan           
}

\newcommand{\Y}{\mathbf{Y}}
\newcommand{\F}{\mathbf{f}}
\newcommand{\Z}{\mathbf{Z}}
\newcommand{\defeq}{\mathrel{\mathop:}=}

\newcommand{\f}{\mathbf{f}}

\newcommand{\rr}{\mathbf{r}}
\newcommand{\W}{\mathbf{W}}

\newcommand{\y}{\mathbf{y}}
\newcommand{\z}{\mathbf{z}}
\usepackage{setspace}
\usepackage[left = 1in, right = 1in, top = 1in, bottom = 1in]{geometry}

\allowdisplaybreaks[4]
\usepackage{mwe}
\usepackage{tikz}
\newsavebox{\measurebox}
\usetikzlibrary{shapes.misc}
\usetikzlibrary{calc,intersections}
\usepackage{pgfplots}
\usetikzlibrary{shapes.geometric}
\pgfplotsset{every axis/.append style={tick label style={/pgf/number format/fixed},font=\scriptsize,ylabel near ticks,xlabel near ticks,grid=major}}
\tikzset{%
  every neuron/.style={
    circle,
    draw,
    minimum size=0.5cm
  },
  rec/.style={
    rectangle,
    draw,
    minimum size=0.5cm
  },
  neuron missing/.style={
    draw=none, 
    scale=2,
    text height=0.333cm,
    execute at begin node=\color{black}$\vdots$
  },
}

\title{\Large \bf Deep Learning in Characteristics-Sorted Factor Models}

\author{ \normalsize  Guanhao Feng\\
\and \normalsize  Jingyu He\\
\and \normalsize  Nicholas G. Polson \\
\and \normalsize  Jianeng Xu\thanks{\scriptsize{We appreciate insightful comments from Will Cong, Serge Darolles, Victor DeMiguel, Li Deng, Jin-Chuan Duan, Shihao Gu (discussant), Bryan Kelly, Soohun Kim (discussant), Markus Pelger, Weidong Tian (discussant), Dacheng Xiu, and Chu Zhang. We are also grateful to helpful comments from the seminar and conference participants at Boston University, CUHK, CityU HK, Jinan University, SHUFE, SUSTech, ESSEC Business School, China International Conference in Finance, Bloomberg, CQAsia Conference, EcoSta, Informs Annual Conference, PanAgora, SOFIE Annual Conference, Schroders, Wolfe Research Conference, Unigestion Factor Investing Conference, Autumn Seminar of Inquire Europe, Australian Finance\&Banking Conference, and New Zealand Finance Meeting. We acknowledge Unigestion Alternative Risk Premia Research Academy, PanAgora, and INQUIRE Europe research awards. Feng acknowledges the ECS grant (9048131) from Hong Kong Research Grants Council and the support from the InnoHK initiative and the Laboratory for AI-Powered Financial Technologies. Feng (Email: \texttt{gavin.feng@cityu.edu.hk}) and He (Email: \texttt{jingyuhe@cityu.edu.hk}) are at the City University of Hong Kong; Polson (Email: \texttt{ngp@chicagobooth.edu}) and Xu (Email: \texttt{jianeng.xu@chicagobooth.edu}) are at University of Chicago.}}
}

\date{\normalsize June, 2023}

\begin{document}
\maketitle

\vspace{-1cm}

\begin{abstract}
\noindent 
This paper presents an augmented deep factor model that generates latent factors for cross-sectional asset pricing. 
The conventional security sorting on firm characteristics for constructing long-short factor portfolio weights is nonlinear modeling, while factors are treated as inputs in linear models.
We provide a structural deep learning framework to generalize the complete mechanism for fitting cross-sectional returns by firm characteristics through generating risk factors -- hidden layers. 
Our model has an economic-guided objective function that minimizes aggregated realized pricing errors. 
Empirical results on high-dimensional characteristics demonstrate robust asset pricing performance and strong investment improvements by identifying important raw characteristic sources.       
	
	
\end{abstract}

\newpage

\section{Introduction}

The Intertemporal Capital Asset Pricing Model \citep[ICAPM,][]{merton1973intertemporal} suggests that a combination of common risk factors can explain the cross section of expected returns. According to the ICAPM, the regression intercept of an asset's return time series on common factors should be zero. 
In contrast to the objective function commonly used in statistics and machine learning, the model's fitness in asset pricing is not determined by the extent of explained variation, but rather by the magnitude of intercepts, or \emph{alphas}. 
Adding factors may lead to statistical overfitting (increasing regression $R^2$), but it does not necessarily improve the asset pricing fitness (decreasing regression intercept). 
This non-arbitrage restriction on alphas implies that practitioners must be cautious when adding factors into their models, as it may not necessarily improve their economic fitness.

A conventional approach to constructing risk factors involves sorting securities on firm characteristics and creating long-short portfolios as proxies for common risk factors. 
Sorted by these risk proxies, higher-risk stocks are supposed to deliver higher expected returns than lower-risk ones. 
For example, the celebrated Fama-French three-factor model \citep{fama1993common} augments the market factor of the capital asset pricing model (CAPM) with SMB (small-minus-big) and HML (high-minus-low).
However, almost all proposed factor models have rejected the zero-alpha hypothesis.
Furthermore, security sorting is vulnerable to the curse of dimensionality \citep{cochrane2011presidential}. Researchers may encounter regions with no stocks, even when using ad hoc double- or triple-sorting methods.

This paper proposes a novel structural deep neural network to approximate the conventional \emph{security sorting}. This method effectively reduces the high dimension of characteristics to a single deep characteristic. 
While deep learning is often associated with forecasting tasks, our approach presents an alternative application by generating latent factors.
Other structural deep learning methods developed for asset pricing include \cite*{gu2021autoencoder}, \cite*{chen2019deep}, and \cite*{fan2022structural}.
We sort trained characteristics, create long-short factors, and estimate factor models to minimize realized pricing errors in a unified framework. 
Notably, the conventional \emph{sorting} securities can be seen as a nonlinear activation function under the deep learning framework. Unlike standard deep/machine learning or statistical models primarily focusing on optimizing statistical fitness, our deep learning optimization prioritizes an economic-guided objective function that reduces unexplained pricing errors within the factor model. This approach provides a promising alternative to the traditional methods of constructing risk factors in empirical asset pricing.

The conventional Fama-French-type characteristics-sorted factors can be conceptualized as an augmented deep factor model. The connections are as follows:
(i) \emph{Inputs} are firm characteristics. The neural network starts from sorting securities on firm characteristics, a nonlinear activation to create long-short portfolio weights.
(ii) \emph{Intermediate features} are risk factors. The factors are linear activations (long-short portfolio weights) on realized returns from the characteristics-sorted direction. 
(iii) \emph{Outputs} are model-implied returns. Reaching the non-arbitrage objective is equivalent to minimizing pricing errors for fitting the factor model to security returns.
The characteristics-sorted factor model can be shown as a ``shallow" learning model for discovering useful characteristics. A motivating example of how momentum factors are dissembled as feed-forward deep learning is introduced in Section \ref{sec: momentum}.

This paper proposes a novel approach to constructing latent factors for asset pricing using deep learning architecture, which differs from existing risk factor literature focusing on selecting or testing observable factors.
For latent factor modeling in asset pricing, the literature includes various principal component analysis (PCA) \citep*{kelly2019characteristics,lettau2020factors,kim2021arbitrage}, deep learning methods \citep{chen2019deep,gu2021autoencoder}, and panel tree \citep*{cong2021asset}.
However, these recent studies have mainly focused on modeling intermediate features (risk factors or basis portfolios) and outputs (asset returns) and skipping the first half of channels.
Our deep learning architecture fills in the missing piece and illustrates the complete mechanism from inputs to immediate features and outputs. 
Under the economic-guided objective, we adopt a structural neural network that includes security sorting, factor generation, and fitting the cross section of stock returns.

The protocol in asset pricing literature is to perform the GRS-type statistical test \citep*{gibbons1989test} on the proposed factor model and stop if the null hypothesis is rejected.  
However, the zero-alpha hypothesis is always rejected, and this approach does not answer the question of the minimum pricing error achievable with the available input data.
In contrast to the hypothesis testing framework, our augmented deep factor model approaches this procedure from an optimization perspective by searching for the optimal set of parameters so that the corresponding deep factors minimize the pricing error. 
In specific, we present a feed-forward neural network consisting of an \textit{input layer} of firm characteristics, \textit{hidden layers} of deep characteristics and risk factors, and an \textit{output layer} of model-implied returns (see Figure \ref{fig:DL_ff5} for the Fama-French example). 
The automatic factor generation process is iterative and receives training feedback through backward propagation, which addresses how much aggregated realized pricing errors can be reduced by optimizing over model parameters at each iteration.

On the methodological side, our approach connects state-of-the-art deep learning optimization with latent factor models in asset pricing. 
Deep learning is a powerful technique with superior pattern-matching capabilities and a flexible yet mysterious ``black box" structure. 
This paper aims to develop a transparent ``white box" deep learning architecture that dissects the characteristics-sorted factor model. 
We propose a systematic approach to automate the security sorting process using deep learning terminologies such as inputs, intermediate features, outputs, and objective functions. 
The structural deep learning approach can complement human learning research by discovering firm characteristics for the fundamental relationship between risk and compensation in a data-driven way. 
Our novel structural augmented deep factor model can better summarize information from high-dimensional characteristics to learn the relationship. 
This model can dissect the expanding zoo of characteristics in asset pricing for a dimension reduction directly on characteristics. Financial economists have been working with the routine procedure of security sorting for decades, and this model can streamline the process.

On the economic side, long-short factors are widely adopted since they (i) reflect compensation for exposure to the risk proxied by underlying characteristics, (ii) can be evaluated as traded portfolios, and (iii) reduce dimension from individual stocks.
However, many of these characteristics' construction formulas are highly similar in accounting, trading, and behavioral perspectives, making them highly correlated. 
The impact of minor differences in formula construction on security sorting, long-short factor performance, and model fitness remains unresolved. 
Previous research has focused on factors or basis portfolios [\textit{intermediate features}] and security returns [\textit{outputs}], neglecting to summarize raw inputs of firm characteristics. 
This paper develops a structural deep learning approach to address this gap to investigate the complete channel from characteristics to risk factors and model-implied returns.

In the empirical analysis, we study monthly data of U.S. individual stocks and test portfolios over the past 50 years. 
The augmented deep factor model has shown comparable statistical and economic model fitness, showing robust improvement for individual stocks or test portfolios. 
The MVE portfolio of deep factors and benchmark factors (market or Fama-French five factors) provide incredibly high  Sharpe ratios (4.57 and 6.49 for in-sample from 1972 to 2011, and 2.49 and 3.00 for out-of-sample from 2012 to 2021) for selected 2-layer DL-CAPM and 3-layer DL-FF5 augmented models. 
We also find the MVE portfolio of constructed deep factors offers perfect hedging performance on the benchmark models. 
Finally, to interpret our deep characteristics from a dimension reduction perspective, we find consistent linear and nonlinear exposures of raw characteristics sources, such as earning surprise, book-to-market ratio, and zero trades.

\subsection{Motivation of Feed-Forward Deep Learning } \label{sec: momentum}

\begin{figure}[h]
	\caption{\textbf{Sorting Securities and Generating Factors}}\label{fig:DL_sort}
	\noindent \footnotesize{This figure provides the feed-forward deep learning procedure for calculating characteristics and creating factors. We start with past returns as raw inputs, and then calculate characteristics as the new inputs used for security sorting. The last layer is the factor generation on long-short portfolio weights obtained from the previous layer plus individual stock returns.}

\vspace{-0.5cm}
	
	\begin{center}	
		\includegraphics[width=0.9\textwidth]{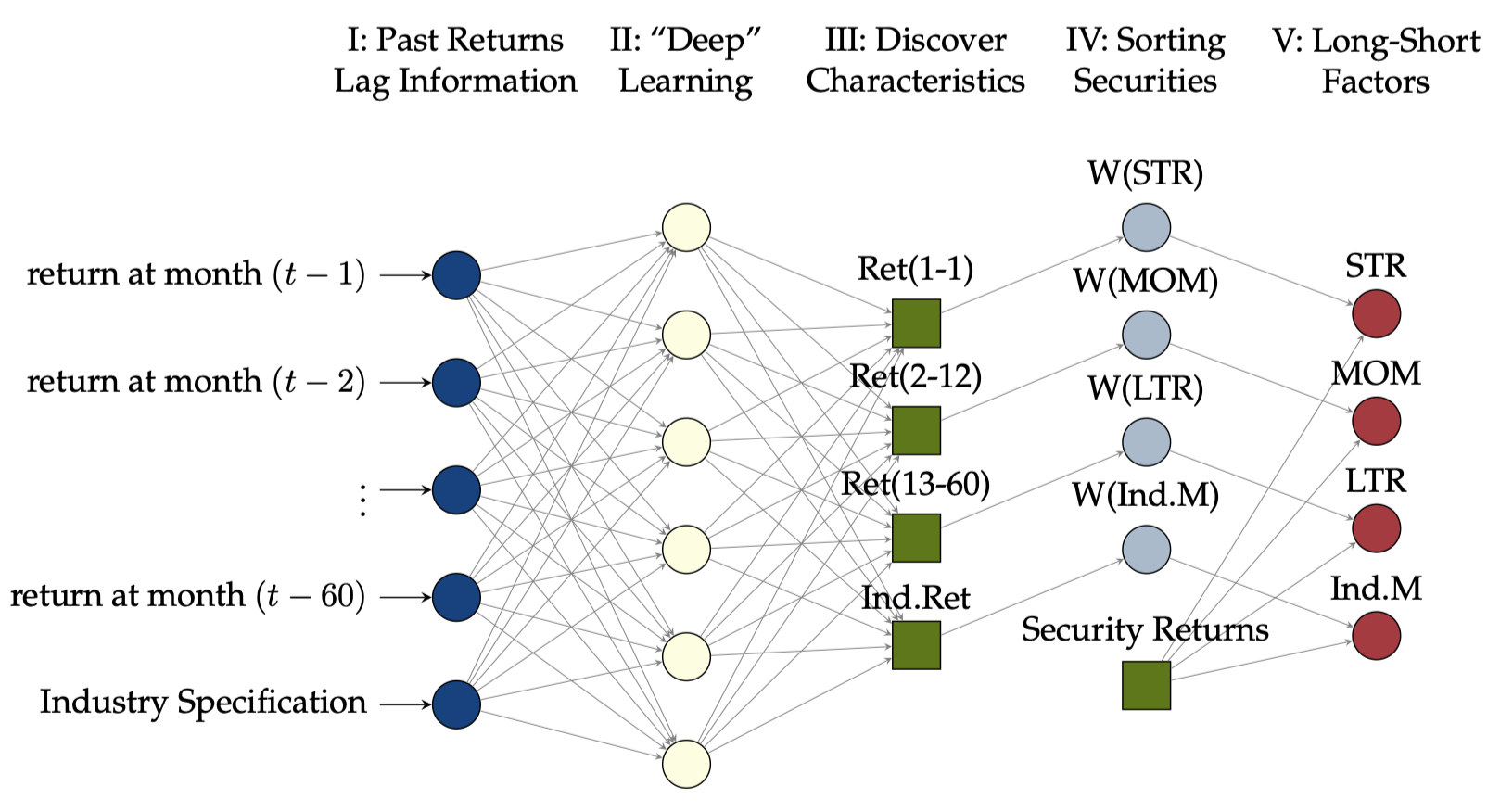}
	\end{center}
	
\vspace{-0.8cm}

\end{figure}

The factor zoo includes many similar firm characteristics to capture the relationship between risk and compensation. Value investing measures such as book-to-market ratio, dividend yield, earning-to-price ratio, and cash flow-to-price ratio are among the proposed measures. However, sorting securities based on these related characteristics can be a tedious trial-and-error process to find the proxy with the best in-sample performance for the specific test assets within the given test period. Before modeling high-dimensional characteristics systematically, a formal description of security sorting is necessary. Figure \ref{fig:DL_sort} depicts a neural network section (without output and loss function definitions) that describes sorting characteristics and generates sorted portfolios.

The standard security sorting approach by ``human learning" can be represented as a nonlinear activation as part of the feed-forward deep learning. 
Let us take the momentum or reversal factor as an example to illustrate this connection. Popular momentum factors include momentum \citep[MOM,][]{carhart1997persistence}, 
long-term reversal \citep[LTR,][]{de1985does}, 
short-term reversal \citep[STR,][]{jegadeesh1993returns},
industry momentum \citep[Ind.M, ][]{moskowitz1999industries}, and others. 
These underlying firm characteristics are essentially (weighted) sums of past returns.
The blue circles in Figure \ref{fig:DL_sort} layer I represent raw inputs of past returns.

In finance research, numerous trial-and-error experiments (e.g., determining the past periods for calculating the past cumulative returns) might be involved in layer II to ``discover" useful characteristics (in layer III). 
This trial-and-error ``human learning" by researchers can be viewed as a multi-layer deep learning process.  
Researchers then sort individual stocks according to their characteristics and generate the long-short portfolio weights, where security sorting is the nonlinear activation. 
The feed-forward deep learning philosophy has been adopted in empirical asset pricing implicitly for a long time but is implemented manually.
The enlarging zoo of similar firm characteristics demands a systematic method to replace layer II with real data-driven ``machine learning".

\subsection{Literature Position}\label{sec: literature}
This paper joins several strands of the empirical asset pricing literature, and the most related one is developing latent factor models for pricing cross-sectional returns. The popular principal components analysis (PCA) has been widely adopted as a statistical factor model. 
The PPCA (Projected PCA) of \cite{kim2021arbitrage} and the IPCA (Instrumental PCA) of \cite{kelly2019characteristics} use firm characteristics as instruments for modeling the time-varying factor loadings and estimating principal components. \cite{kelly2020modeling} find that IPCA also works well for corporate bond returns. 
\cite{lettau2018estimating,lettau2020factors} provide a regularized estimation for risk premia, RP-PCA (Risk Premia PCA), identifying factors with small time series variance but are useful in the cross section. 
\cite{kozak2018interpreting} show that PCA of anomaly portfolios works as well as a reduced-form factor model in explaining anomaly portfolios. 
Like IPCA, \cite{gu2021autoencoder} incorporate one deep learning technique, auto-encoder, to construct nonlinear components. 
\cite{chen2019deep} adopt the generative adversarial network to estimate the SDF model that explains all asset returns from the conditional moment constraints implied by no-arbitrage, and \cite{cong2021asset} generates the SDF model on the basis portfolio generated by the panel tree.

A critical difference of our latent factor generation is that it does not require the \textit{intermediate features} of basis portfolios but directly approximates the long-short factor on individual stocks.
We compare the results of our method with IPCA and RP-PCA in the empirical analysis. Our deep learning framework differs from the standard PCA in several ways:
(i) The dimension reduction is applied directly to firm characteristics [inputs] rather than factors or basis portfolios [intermediate features], or security returns [outputs]. 
(ii) We conduct dimension reduction through deep learning architecture that can learn both nonlinearity and interactions of inputs, while PCA can only extract linear components. 
(iii) PCA requires a balanced panel data structure, whereas ours allows for an unbalanced panel of individual stock returns.

Our augmented deep factor model effectively connects the literature on statistical factor models with the subsequent literature on economic factor models using characteristics-sorted factors.
However, the existing literature primarily focuses on modeling between inputs (factors) and outputs (asset returns) and skips the intermediate channel: \textit{how these risk factors are generated}. 
We fill in the missing pieces with our bottom-up approach, from firm-level characteristics to latent factors for fitting cross-sectional returns.

Furthermore, this paper joins the recent literature on selecting and testing the zoo of risk factors or characteristics for the cross section of stock returns.
A seminal paper of \cite*{harvey2016and} raise a multiple testing issue to challenge 300 factors discovered in the literature,
\cite*{green2017characteristics} find only 12 out of 94 characteristics are reliably independent determinants in non-microcap stocks,
\cite*{hou2017replicating} and \cite*{jensen2022there} revisit the replication of the factor zoo for the U.S. and global market.
\cite*{light2017aggregation} use partial least squares (PLS) to aggregate information on firm characteristics. 
\cite*{demiguel2020transaction} offer the economic rationale of why many characteristics are needed in investing portfolios by considering transaction cost.

Our variable selection, dimension reduction, and nonlinear transformation are implemented in the characteristics space.
\cite*{kozak2020shrinking} use a shrinkage estimator on the SDF coefficients for characteristic-based factors.
\cite*{feng2020taming} develop a regularized two-pass cross-sectional regression to evaluate the incremental contribution of risk factors.
The recent development for variable selection methods to firm characteristics or risk factors include \cite*{chinco2021estimating}, \cite*{dong2022anomalies}, and \cite*{bryzgalova2021bayesian}.
From a perspective of uncommon factors, \cite*{cong2022uncommon} develop a Bayesian clustering model for a joint problem of factor selection and asset clustering.

Finally, our paper contributes to the wide and successful applications of machine learning and deep learning in finance. \cite*{heaton2017deep} introduces deep learning decision models for financial prediction and classification problems. \cite{gu2020empirical} provide a comprehensive empirical investigation of return prediction performance using firm characteristics by multiple machine learning algorithms. 
Other notable new methods for predicting stock returns include adaptive group LASSO \citep*{freyberger2020dissecting} and forecast combination \citep*{guofu2018characteristics}.
These works focus on characteristics and security returns but neglect the economic structure of risk factors, while our paper fills in this hidden piece. 
Alternative predictors of stock returns include text data \citep*{ke2019predicting} and image data \citep*{jiang2022re}. 
Machine learning return predictability to other assets include treasury bonds \citep*{bianchi2021bond,fan2022real} and corporate bonds \citep*{bali2021ml,he2021predicting}. 
Finally, for portfolio optimization studies via machine learning, \cite*{cong2020alphaportfolio} provide a reinforcement learning approach, \cite*{feng2022factor} consider the Bayesian Hierarchical modeling, and \cite*{bryzgalova2020forest} develop regularized portfolios using basis portfolios generated by the decision tree.

The rest of the paper is organized as follows: Section \ref{sec: architecture} illustrates why the characteristics-sorted factor can be conceptualized in deep learning. 
Section \ref{sec: algo} provides technical details for our structural deep learning framework.  
Section \ref{sec: empirical} shows the empirical findings and comparison with other PCA models.  
Section \ref{sec: summary} concludes the paper, and additional implementation details are provided in Appendix \ref{app: optim}.

\section{Deep Learning and Empirical Asset Pricing}\label{sec: architecture}
Section \ref{sec: pricing} presents the asset pricing optimization on realized pricing errors. 
Section \ref{sec: char-model} presents a deep learning reformulation of the characteristics-sorted factor model, while Section \ref{sec: FF intro} refers to implementing Fama-French models as a special case.

\subsection{Asset Pricing Optimization}\label{sec: pricing}
The standard asset pricing linear factor model decomposes excess asset returns on factor risk premia as follows:
\vspace{-0.5cm}
\begin{equation}\label{eqn: er1}
	E_{t-1}\left(r_{i,t}\right) = \boldsymbol{\beta}_{i,t-1}^\intercal\boldsymbol{\lambda}_\f,
\end{equation}
where $r_{i,t}$ is the excess return of asset $i$ at time $t$, $\boldsymbol{\beta}_{i,t-1}$ is the factor loading updated at time $t-1$, and $\boldsymbol{\lambda}_\f$ is the risk premia of the $K \times 1$ factors $\f_t$. If $\f_t$ is tradable, a commonly used estimator for $\boldsymbol{\lambda}_\f$ is the factor past average return. Any potential mispricing $\alpha_{i}$ is added on the right-hand side. 
\vspace{-0.5cm}
\begin{equation}\label{eqn: er2}
	E_{t-1}\left(r_{i,t}\right) = \alpha_{i} + \boldsymbol{\beta}_{i,t-1}^\intercal\boldsymbol{\lambda}_\f.
\end{equation}

The asset pricing literature focuses on $\alpha_i$, the intercept of time series regression for mispricing. To test an asset pricing model, a joint test on the vector of ${\alpha_i}$ across assets is suggested by \cite{gibbons1989test}. The GRS test statistic's kernel is a weighted sum of quadratic alphas, $\boldsymbol{\alpha}^\intercal \Sigma^{-1}{\boldsymbol{\alpha}}$, where $\Sigma$ is the covariance matrix across assets. 
Cross-sectional variations adjust the weight of pricing errors. If a sufficient factor model exists, the quadratic term of pricing errors should be statistically and economically insignificant. However, no such factor model has been found in the academic literature.

This paper proposes an alternative approach to addressing the problem of whether a factor model is sufficient to explain the cross section. Rather than solely testing the adequacy of a factor model, we suggest generating additional latent factors to minimize aggregate realized pricing errors. 
While machine learning and deep learning methods are often criticized for overfitting, overfitting only improves the in-sample $R^2$ without necessarily reducing the regression intercept or $\alpha_i$. 
Our model aims to improve asset pricing models by reducing aggregated realized pricing errors, rather than solely improving the statistical fitness of the model. 
To achieve this, they include pricing errors in the objective function and adopt a characteristics-sorted factor model within the deep learning framework.

Furthermore, \citet{cochrane2011presidential} asserts that return-factor covariances are stable functions of asset characteristics. Similar to \cite{gu2021autoencoder}, we specifically train a nonlinear neural network of betas on lagged characteristics. Therefore, the population model in Equation (\ref{eqn: er2}) is 
\vspace{-0.5cm}
	\begin{equation}\label{eqn: reg_interaction}
	\begin{aligned}
		r_{i,t} &= \alpha_{i} + \boldsymbol{\beta}_{i,t-1}^\intercal \f_t + \epsilon_{i,t},\\
		\boldsymbol{\beta}_{i,t-1} &= G\left(\z_{i,t-1}\right)
	\end{aligned}
	\end{equation}
where the parallel neural network of $G(\cdot)$ for betas are learned jointly with the latent factor $\f_t$, see Section \ref{sec: deep f} for the detailed functional form. 

Let $\widehat r_{i,t}$ denote a linear portfolio constructed with traded factors $\f_t$ to mimic the asset return $r_{i,t}$. Since factors and their interaction with characteristics are traded in the time series regression, $\widehat r_{i,t} = \boldsymbol{\beta}_{i,t-1}^\intercal \f_t$ is a linear combination of portfolios \textit{without an intercept}. The intercept of $\alpha_i$ equals to the average fitting difference $\frac{1}{T}\sum^T_{t=1} \alpha_{i,t}$ if the residual average is zero, where $\alpha_{i,t} = r_{i,t} - \widehat r_{i,t}$.

The objective function in our asset pricing optimization is determined by the property of the traded portfolio for alphas. The GRS test statistic kernel, with an equal weighting of $\sum^{N}_{i=1} \alpha_i^2$, is used to measure the aggregate average pricing errors across assets. 
It is important to note that our aggregate \textit{realized pricing errors} is expressed as  $\sum^{N}_{i=1} \sum^T_{t=1} \alpha_{i,t}^2 = \sum^{N}_{i=1} \sum^T_{t=1} (\alpha_{i}+\epsilon_{i,t})^2$, which includes both average pricing errors and residual errors, as shown in Equation (\ref{eqn: variation}). 
Reducing aggregate realized pricing errors might explain variations of average returns in the cross section (average pricing errors) and time series (residual errors). 
This paper contributes to the growing body of literature, including recent studies such as \cite{chen2019deep} and \cite{cong2021asset}, that generate latent factors to minimize average or realized pricing errors. 
However, it is worth noting that the weight for cross-sectional variation is typically smaller than that for time-series variation.

\subsection{Characteristics-Sorted Factors in Deep Learning}\label{sec: char-model}
To model risk-free adjusted excess returns, our augmented deep factor model generates $P_d$ deep factors $\f_{d,t}$ in addition to a $P_b$-factor benchmark model $\f_{b,t}$, which can be CAPM or Fama-French 5 factor models, to price individual stock returns jointly. We use $\f_t = [\f_{d,t}^\intercal, \f_{b,t}^\intercal]^\intercal$ denoting the set of both deep and benchmark factors.\footnote{We thank an anonymous referee for suggesting fitting individual stocks on a conditional factor model. The previous version of the paper considers an indirect objective of fitting portfolio returns.} 
We form the realized return predictor $\widehat r_{i,t}$ as a linear combination of $\f_{d,t}$ and $\f_{b,t}$ without an intercept. Therefore, the fitting error, $\alpha_{i,t}$, measures the cross-sectional and time-series variation. The conventional pricing error is the average fitting difference $\frac{1}{T}\sum^T_{t=1} \alpha_{i,t}$. The full model is 
\vspace{-0.5cm}
\begin{equation}\label{eqn: return-factor}
	\begin{aligned}
		\widehat r_{i,t} & = \boldsymbol{\beta}\left(\z_{i,t-1}\right)^\intercal \f_t \\
		                 & = \boldsymbol{\beta}_d\left(\z_{i,t-1}\right)^\intercal \f_{d,t} + \boldsymbol{\beta}_b\left(\z_{i,t-1}\right)^\intercal \f_{b,t}, \\
		\alpha_{i,t}     & = r_{i,t} - \widehat r_{i,t},                                                                     \\
		\f_{d,t}             & = \W_{t-1}\rr_t,                                                                                  \\
		\W_{t-1}         & = H\left(\Z_{t-1}\right),\\
  		\boldsymbol{\beta}\left(\z_{i,t-1}\right)&= G\left(\z_{i,t-1}\right)
,                                                                                    
	\end{aligned}
\end{equation}
where $\boldsymbol{\beta}(\z_{i,t-1})^\intercal = [\boldsymbol{\beta}_d(\z_{i,t-1})^\intercal, \boldsymbol{\beta}_b(\z_{i,t-1})^\intercal]$ are the dynamic betas corresponding to the latent deep factors and benchmark factors, respectively. Furthermore, $\rr_t = (r_{1,t},\cdots, r_{N,t})$ denote a vector of $N$ individual firm returns at month $t$, and $\W_{t-1}$ is a $P_d \times N$ matrix for long-short portfolio weights determined at month $t-1$. $\Z_{t-1}= (\z_{1,t-1}, \cdots, \z_{N, t-1})$ is a $K_0\times N$ dimension matrix of lagged firm characteristics.

The deep factors $\f_{d,t}$ are long-short portfolios constructed by sorting individual firms on deep characteristics generated from lagged firm characteristics $\Z_{t-1}$ with dimension $K_0 \times N$. 
The nonlinear function $H(\cdot)$ transforms stock characteristics into long-short portfolio weights. 
It is flexible for different purposes, given the different structures of the underlying network. For example, suppose the network reduces to zero layers and outputs $K_0\times N$ dimensional $\W_{t-1}$. In that case, it corresponds to the characteristics-managed portfolio $\Z_{t-1}\rr_t$ in \cite{kelly2019characteristics}, which uses the normalized $\Z_{t-1}$ to form $\W_{t-1}$.
The network can also be designed to provide a univariate sort and create long-short directions $\{1, 0, -1\}$ on each characteristic, plus the multiplication with equal or value weights to form $\W_{t-1}$. 
The architecture of $H(\cdot)$ is discussed further in Section \ref{sec: algo}. Furthermore, the dynamic beta $\boldsymbol{\beta}\left(\z_{i,t-1}\right)$ is driven by firm characteristics, modeled by a separate neural network but trained jointly with $H(\cdot)$, see Section \ref{sec: deep f} for details.

With the notation $\{\f_{d,t}, \rr_t, \W_{t-1}, \Z_{t-1}\}$, the characteristics-sorted factor model is transparently formulated in the above deep learning architecture. 
First, the inputs are lagged characteristics $\Z_{t-1}$. 
Second, sorting securities in the month $t-1$ according to the deep characteristics, we obtain intermediary features $\W_{t-1}$.
Third, latent factor $\f_{d,t}$ is a traded portfolio of individual stock realized returns $\rr_t$.
Forth, functions for factor loadings, $\boldsymbol{\beta}(\cdot)$ are conditional on characteristics $\z_{i,t-1}$ and learned jointly with the deep factors.
Finally, the deep factors can be constructed to complement a benchmark model $\{\f_{b,t}\}$ to fit $\rr_t$.

The characteristics-sorted factor model above has prediction power since the portfolio weights only use information lagged in one period. The deep learning latent factor $\f_{d,t}$ is built with long-short portfolio weights at month $t-1$ and individual firm returns at month $t$. 
This factor model return differs from those models in \cite{freyberger2020dissecting} and \cite{gu2020empirical} for predicting returns via firm characteristics, because we use realized returns $\{\rr_t, \f_{b,t}\}$ as second and third inputs.

The above augmented deep factor model generates $\f_{d,t}$ while controlling for the benchmark factors $\f_{b,t}$ (such as CAPM or FF5), which is consistent with the conventional practice that researchers admit new factors based on incremental signals. 
The objective is to minimize the quadratic sum of realized pricing errors aggregated over all periods and assets:
\vspace{-0.2cm}
\begin{equation}\label{eqn: loss}
	\begin{aligned}
		\mathcal{L}_\lambda & = \sum_{i=1}^{N} \sum_{t=1}^{T_i} \left(r_{i,t} - \boldsymbol{\beta}\left( \z_{i, t-1}\right)^\intercal \f_t\right)^2 + \lambda \times \text{penalty(DL)}, 
	\end{aligned}
\end{equation}
where $\lambda$ is a tuning parameter that regularizes parameters trained in the neural network, which helps to stabilize the network model training and introduces sparsity for selecting characteristics.
In our empirical study, we perform two-fold deterministic cross-validation to select $\lambda$ from a sequence of candidates.

\subsection{Fama-French Five-Factor Model in Deep Learning}\label{sec: FF intro}
\begin{figure}[h]
	\caption{\textbf{Fama-French Five-Factor Model in Deep Learning}}\label{fig:DL_ff5}
	\noindent \footnotesize{This figure provides the feed-forward deep learning representation of generating the Fama-French five factors using firm characteristics to calculate the objective function, aggregate pricing errors. Lagged characteristics are inputs, long-short factors are hidden neurons, and factor-model fitted returns are outputs.}

\vspace{-0.4cm}
	
	\begin{center}	
		\includegraphics[width=0.9\textwidth]{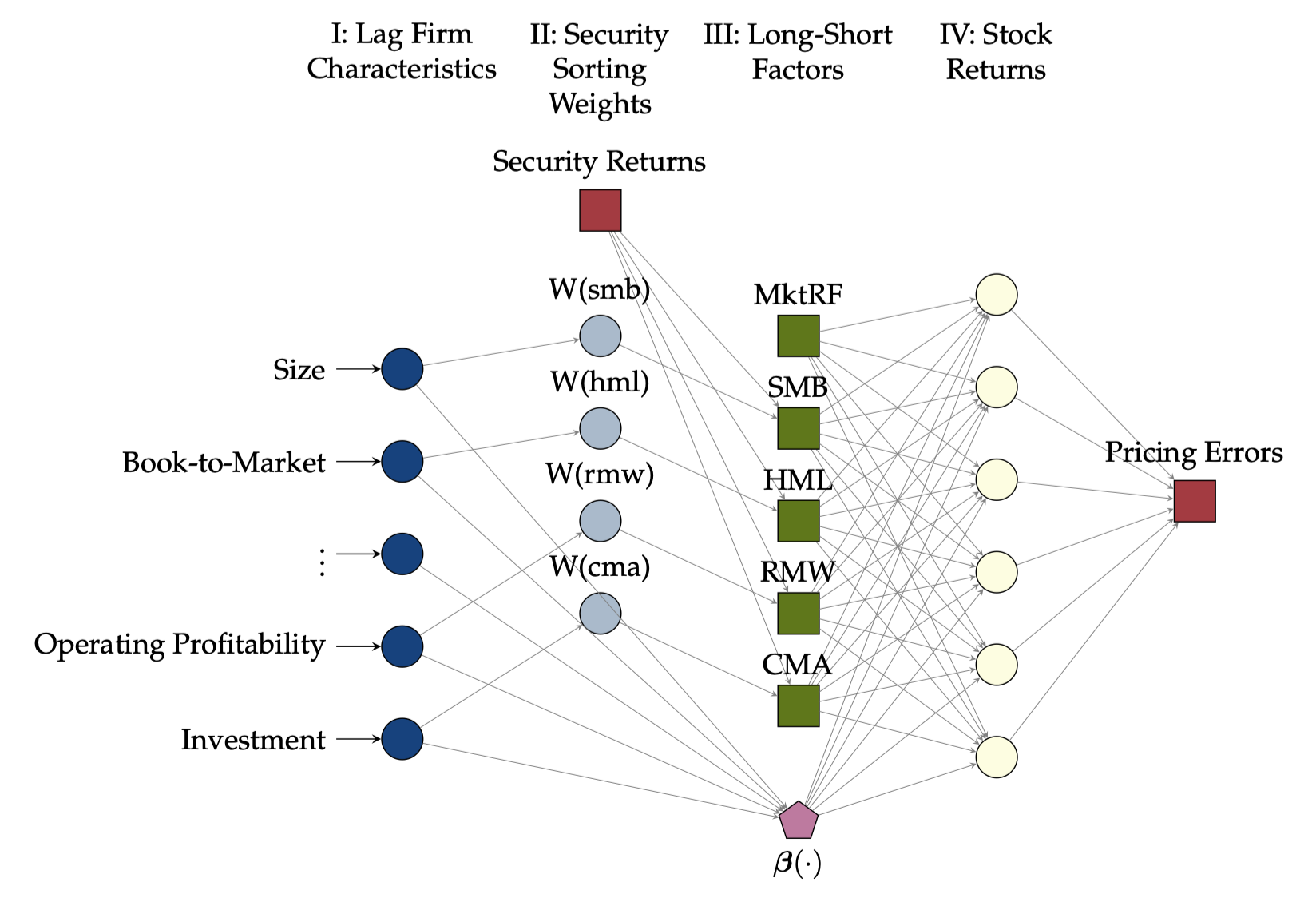}
	\end{center}

\vspace{-0.8cm}

\end{figure}

This subsection reformulates the Fama-French five-factor model (FF5) within a complete deep learning architecture for a characteristics-sorted factor model. Figure \ref{fig:DL_ff5} illustrates the steps from characteristics to security sorting, long-short factors, factor-model fitted returns, and aggregate pricing errors.

Researchers begin by identifying ``discovered" characteristics, represented by blue circles in layer I. The hidden function $H(\cdot)$ used for this ``discovery" is \emph{determined before the blue circles}. 
Then they sort individual firms on lagged characteristics to determine the long-short portfolio weights, as shown in the cyan circles in layer II.\footnote{If a firm has missing values in some periods, it is excluded from the security sorting for these periods. Therefore, security sorting works perfectly for imbalanced panel data with missing values.} 
In layer III, factors are constructed using the portfolio weights from the previous layer and the realized returns of individual stocks as additional inputs. 
Adding the Market factor produces an augmented factor model to explain realized returns of test assets\footnote{Empirical studies typically use characteristics-sorted portfolios as test assets.} in the purple circles. 
Factor loadings, represented by $\boldsymbol{\beta}(\cdot)$, can be considered a function of lagged characteristics in the conditional model or constant values in the unconditional model, as shown by the pink pentagon in layer III. 
Finally, the objective function for minimizing the aggregate average or realized pricing errors is represented by the red rectangle.

The Fama-French five-factor model adds four additional factors $\f_{d,t}$ to CAPM, where $\f_{b,t}$ is the Market factor (MktRF). Those four characteristics (size, book-to-market, operating profitability, and investment) are $\Z_{t-1}$. Stock weight $\W_{t-1}$ is determined by (bivariate-) sorting directions to form long-short factors. \cite{FAMA2015} have tested these four additional factors over CAPM to confirm their significance. The four additional factors in our deep learning diagram are trained by controlling the CAPM benchmark model to minimize the objective function of aggregate pricing errors.

Researchers have been working on nonlinear modeling for a long time because security sorting is a nonlinear function to determine long-short portfolio weights.
Anomalies are typically calculated using pre-tested formulas to assist cross-sectional pricing. Still, the drawback is that the usefulness of a characteristic (or an anomaly) is tested \emph{ex-post} statistically, and feedback on model fitting never returns to characteristics' construction procedures.
A sequential optimization approach is proposed to address this issue. 
The automatic factor generation receives training feedback through backward propagation, optimizing model parameters to minimize pricing errors. By refitting the model sequentially and optimizing the formula of ``deep characteristics" based on feedback from the loss function, our approach offers an alternative to the one-time hypothesis testing protocol.

\section{Architecture of Deep Neural Network}\label{sec: algo}

\begin{figure}[h]
	\caption{\textbf{Deep Learning Network Architecture}} \label{fig:network}
	\noindent \footnotesize This figure provides a visualization of deep learning architecture. The firm's characteristics $\Z_{t-1}$ are selected and transformed into deep characteristics $\Y_{t-1}$. Then, we ``sort" $\Y_{t-1}$ to generate factor weight $\W_{t-1}$. The deep factors $\f_{d,t}$ and benchmark factors $\f_{b,t}$ are used to price individual stock returns $\rr_{t}$.
	
\vspace{-0.3cm}	
	
	\begin{center}	
		\includegraphics[width=0.9\textwidth]{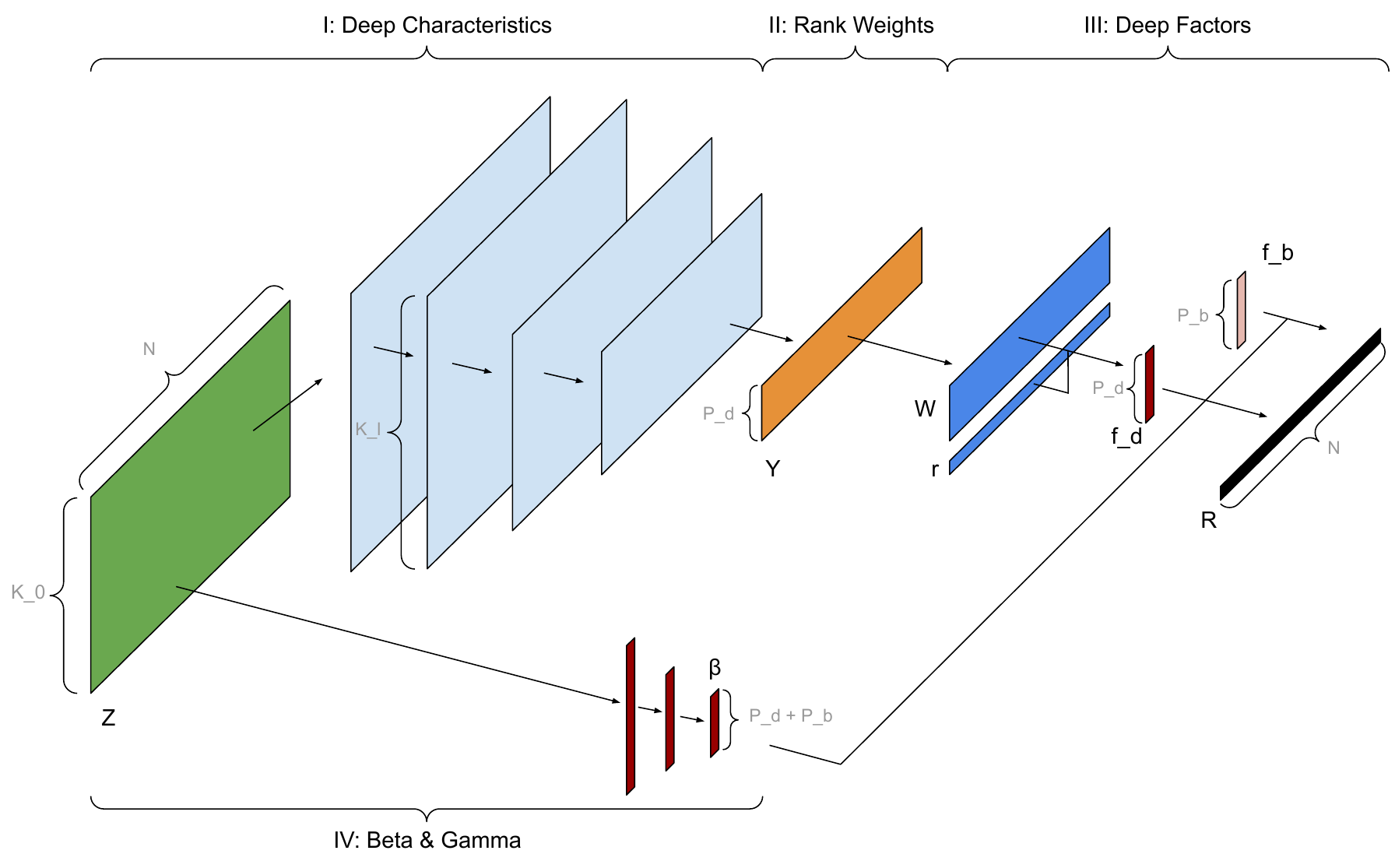}
	\end{center}
	
\vspace{-0.5cm}		
\end{figure}

This section introduces the deep learning architecture (Figure \ref{fig:network}) and its four major parts: I: Deep Characteristics, II: Rank Weights, III: Deep Factors, and IV: Dynamic Factor Loadings. 
In specific, Section \ref{sec: deep char} illustrates how the dimension reduction on the \emph{[inputs]} firm characteristics performed in the feed-forward neural network. Section \ref{sec: weight} calculates long-shot portfolio weights, and Section \ref{sec: deep f} calculates the \emph{[intermediate features]} deep factors. 
Section \ref{sec: loss} describes the optimization objective and summarizes the complete augmented deep factor model. 
We first clarify notations. A typical training observation indexed by time $t$ includes the following types of data: 
\vspace{-0.2cm}
\begin{eqnarray*}
\begin{aligned}
	  & \left\{r_{i,t}\right\}_{i=1}^{N}, \;\text{excess returns of $N$ individual stocks,}                         \\
	  & \left\{z_{k,i,t-1}: 1\leq k \leq K_0\right\}_{j=1}^{M}, \;\text{$K_0$ lagged characteristics of $N$ individual stocks,} \\
	  & \left\{f_{d,t}\right\}_{d=1}^{P_b},  \;\text{$P_b$ benchmark factors}.                                       
\end{aligned}
\end{eqnarray*} 
The bold symbols $\left\{\rr_t, \z_{i, t-1}, \Z_{t-1}, \f_{b,t}\right\}$ represent grouped assets data, where $\rr_t$ is an $N\times 1$ vector of returns, $\z_{i, t-1}$ is a $K_0\times 1$ vector of characteristics of asset $i$, $\Z_{t-1}$ is a $K_0 \times N$ matrix of characteristics for all assets, and $\f_{b,t}$ is a $P_{b} \times 1$ vector of benchmark factors. $\Y_{t-1}$ is a $P_d \times N$ matrix of deep characteristics, and  $\y_{t-1} = (y_{1},\cdots, y_N) $ is a $N\times 1$ vector representing one deep characteristic (a row of $\Y_{t-1}$).  In Section \ref{sec: emp_data}, we have $N = 3,000$ stocks, $K_0 = 60$ characteristics, and $P_{b} = 1 \text{ or } 5$ for CAPM or FF5 as benchmarks. Table \ref{tab: mechanism} below summarizes details of the deep learning structure, and Figure \ref{fig:network} visualizes it.

\begin{table}[h]
	\caption{\textbf{Deep Learning Mechanism for Variable Dimensions}} \label{tab: mechanism}
	\noindent {\footnotesize This table summarizes the augmented deep factor model. The initial inputs are raw characteristics; the final outputs are model-fitted individual stock returns. In each layer, the network takes the output from the immediate previous layer as its input and the additional input if needed. The other inputs include individual stock returns $\rr_t$ for constructing deep factors $\f_{d,t}$ and the benchmark model $\{\f_{b,t}\}$. The last column references the equations in the main text.}
	
	\begin{center}
		\footnotesize{
			\begin{tabular}{@{}ccccccc@{}}
				\toprule
				                       & Dimension                      & Output               & Inputs         & Operation                                                                         & Parameters                     & Equation               \\ \midrule

				Lagged characteristics & $K_0\times N$                  & $\Z_{t-1}^{[0]}$      & $\Z_{t-1}$     & $\Z_{t-1}^{[0]}:= \Z_{t-1}$                                                        &                             & (\ref{eqn: deepchar})  \\
				& $\vdots$            & $\vdots$                                               &                & $\vdots$                                                                        & $\vdots$                                                \\
				                       & $K_l \times M$                 & $\Z_{t-1}^{[l]}$      &                & $F^{[l]} \left(\Z_{t-1}^{[l-1]}\right)$                                            & $\left(A^{[l]},b^{[l]}\right)$ & (\ref{eqn: deepchar})  \\
				& $\vdots$            & $\vdots$                                               &                & $\vdots$                                                                        & $\vdots$                                                \\
				Deep Characteristics   & $K_L \times N$                 & $\Y_{t-1}$            &                & $F^{[L]} \left(\Z_{t-1}^{[L-1]}\right)$                                            & $\left(A^{[L]},b^{[L]}\right)$ & (\ref{eqn: deepchar})  \\
				\\
				Rank Weights           & $P_d \times N$                 & $\W_{t-1}$            &                & $h^{[0]}\left(\y\right)$                                                          &                                & (\ref{eqn: weight})    \\
				\\
				Deep Factors           & $P_d \times 1$                 & $\f_{d,t}$           & $\mathbf{r}_t$ & $\W_{t-1}\mathbf{r}_t$                                                            &                                & (\ref{eqn: weight-DL}) \\
				\\
				Dynamic Loadings       & $\left(P_d+P_b\right)\times 1$ & $\boldsymbol{\beta}$ & $\Z_{t-1}$     & $\boldsymbol{\beta} = G\left(\Z_{t-1}\right)$                          & $\left(A_G, b_G\right)$        & (\ref{eqn: beta})      \\
				\\
				Security Returns       & $N\times 1$                    & $\widehat r_t$       & $\f_{b,t}$     & $\boldsymbol{\beta}^\intercal \left[\f_{d,t}^\intercal, \f_{b,t}^\intercal\right]^\intercal$ & $\boldsymbol{\beta}$           & (\ref{eqn: returns})   \\
				
				\bottomrule
			\end{tabular}}
	\end{center}

\end{table}

\subsection{Deep Characteristics} \label{sec: deep char}
We first introduce the design of neural networks to generate $P_d$ deep characteristics from the raw inputs, the key to constructing deep factors. This operation is the ``deep" part that learns nonlinearity and interactions of the raw input characteristics, and reduces the dimension from $K_0$ to $P_d$ deep characteristics. 
The architecture is as follows for asset $i=1, 2,\cdots, N$ and $l$-th layer of the neural network (note that the raw input is $l=0$ layer):
\vspace{-0.2cm}
\begin{align*} 
	\z_{i, t-1}^{[l]} & = F^{[l]} \left( A^{[l]} \z_{i, t-1}^{[l-1]} + b^{[l]} \right),\; \text{for } l =1, 2, \cdots, L \\
	\z^{[0]}_{i, t-1} & \defeq \left[z_{1,i, t-1},\cdots,z_{K_0,i, t-1}\right]^\intercal,                                                  
\end{align*} 
where $\z_{i, t-1}^{[l]}$ is the $i$-th column of a $K_l\times N$ matrix $\Z^{[l]}_{t-1}$. We set $K_L = P_d$ representing the output of the final layer corresponding to the number of deep characteristics. The univariate activation function $F(\cdot) = \text{tanh}(\cdot)$ broadcasts to every element of a matrix. The parameters to be trained in this part are deep learning weights $A$ and biases $b$, namely,
\vspace{-0.2cm}
$$
\left\{(A^{[l]},b^{[l]}): A^{[l]}\in \mathbb{R}^{K_l\times K_{l-1}}, b^{[l]} \in \mathbb{R}^{K_l}\right\}_{l=1}^L.
$$

Distinct from the standard deep neural network structure where each neuron is fully connected to all neurons in the previous layer, the transformations are performed column-wise with no communication across different stocks. 
All transformations performed in this part are within each stock. The data (and intermediate results) of two different stocks are separated and do not interfere with each other. This setting allows the deep characteristics of a specific individual stock to be determined by itself but not by other stocks. Then we sort all stocks on the deep characteristics, collecting information across different stocks. 
Our empirical study found that the results are not sensitive to the choice of activation function, and the combination and transformation of characteristics play more important roles. 
This multi-layer structure summarizes information from the raw characteristics to lower dimensional deep characteristics. We later introduce how to create deep factors by sorting on deep characteristics in Section \ref{sec: deep f}. Ultimately, the deep learning algorithm estimates parameters here by optimizing for the loss function \ref{sec: loss}, and learns how to combine and select useful characteristics in a data-driven way.

With some abuse of notation, we express the $P_d \times N$ deep characteristics by $\Y_{t-1}$, the output of the deep learning architecture that generates deep characteristics, 
\vspace{-0.2cm}
\begin{eqnarray} \label{eqn: deepchar}
	\Y_{t-1}       & \defeq & \Z_{t-1}^{[L]},                                                               \\
	\Z_{t-1}^{[l]} & \defeq & F \left( A^{[l]} \Z_{t-1}^{[l-1]} + b^{[l]} \right),\; \text{for } l =1, 2, 3, \cdots, L \\
	\Z_{t-1}^{[0]} & \defeq & \Z_{t-1}.                             
\end{eqnarray}
Unlike the standard feed-forward neural network, the $l$-th layer in our architecture is a neural matrix $\Z_{t-1}^{[l]}$. Each row of $\Z_{t-1}^{[l]}$ is a $N\times 1$ vector representing the $k_l$-th intermediary characteristics for $N$ individual stocks, $k_l = 1, 2, \cdots, K_l$. We explicitly make all the columns (stocks) share the same parameters $A^{[l]}$ and $b^{[l]}$, whose dimensions are independent of $N$. Therefore, the formula for deep firm characteristics is the same for every stock.

\begin{figure}[h]
	\caption{\textbf{Deep Neural Network of $\Z^{[l-1]}_{t-1}\rightarrow \Z_{t-1}^{[l]}\rightarrow \Z_{t-1}^{[l+1]}$. }} \label{fig:deepchar}
	\noindent \footnotesize This figure shows how the feed-forward neural network forwards from $\Z_{t-1}^{[l-1]}$ to $\Z_{t-1}^{[l+1]}$. 
	The lines connecting two layers represent affine transformation, and the circles represent the activation function.

\vspace{-0.3cm}		
	\begin{center}	
		\includegraphics[width=0.7\textwidth]{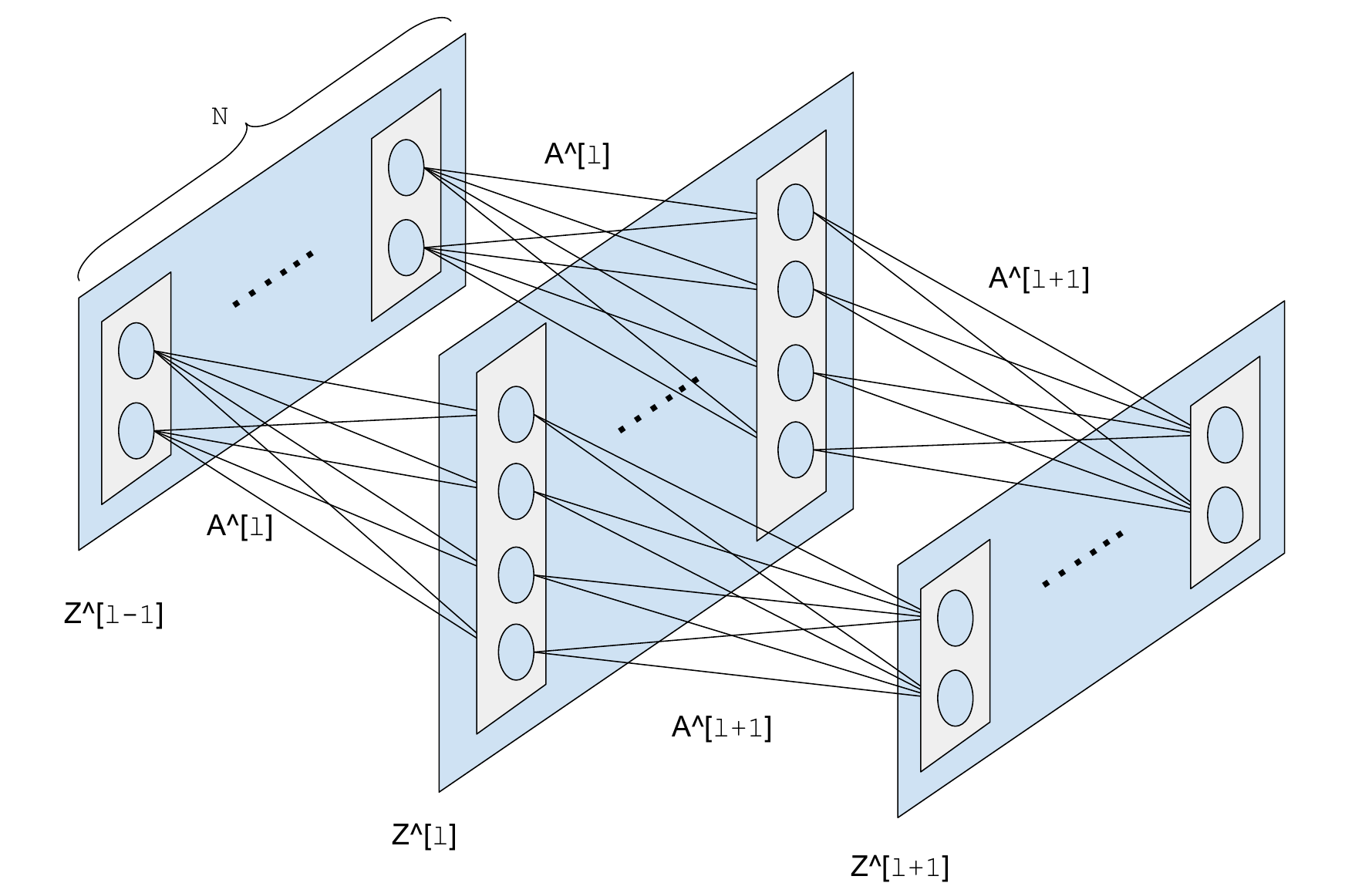}
	\end{center}

\vspace{-0.5cm}		
\end{figure}

Here, $K_l$ denotes the dimension of the $l$-th layer because the number of columns is fixed as $N$ for all $\Z_{t-1}^{[l]}$'s. Figure \ref{fig:deepchar} illustrates how our deep neural network operates by showing a sample architecture from the $(l-1)$-th to the $(l+1)$-th layer, where $K_{l-1}=K_{l+1}=2$ and $K_l=4$. The Fama-French-type factors use $\Y_{t-1}\defeq \Z_{t-1}$ for sorting in the latter part (by hiding hidden layers). By contrast, $\Z_{t-1}^{[0]} \defeq \Z_{t-1}$ in our deep neural network goes through multiple layers of affine transformations and nonlinear activations and ends up with a low-dimensional deep characteristic $\Y_{t-1}$. Here, the layer sizes $\{K_l\}_{l=1}^{L}$, and the number of layers $L$ are architecture parameters chosen by model designers.

\subsection{Sorting and Nonlinear Weights} \label{sec: weight}
Next, we introduce the particular design of the nonlinear activation function to approximate the security sorting in empirical asset pricing. Sorting firm characteristics to generate sorted portfolios and long-short factors is underappreciated nonlinear modeling for linear factor models because researchers usually assume these are given and observable. Our structural deep neural network aims to systematically approximate security sorting and construct long-short factors approximate security sorting and construct long-short factors systematically.

Researchers usually long (and short) top (bottom) 10\% or 20\% of stocks for equal or value weights to create factors. 
However, two recent papers adopt the general rank weights for creating factors. \cite{frazzini2014betting} construct Betting-against-Beta with a ``rank weighting" and assign each stock to either the ``high" portfolio or the ``low" portfolio with a weight proportional to the cross-sectional ranked market beta. \cite{novy2022betting} add a further discussion to compare different portfolio weighting schemes: rank (linear) weights versus equal weights. We follow these studies and incorporate nonlinear rank portfolio weights as activation outputs from deep learning.

Given the deep characteristics $\Y_{t-1}$ summarized by deep learning structure in Equation (\ref{eqn: deepchar}), we construct the portfolio weight $\W_{t-1}$ as follows. We define $h^{[0]}: \mathbb{R}^{M} \rightarrow [-1, 1]^N$ to calculate the portfolio weights based on the rankings of the deep characteristics.  When the variable is a matrix, it broadcasts to all rows. Let $\y_{t-1} = (y_{1},\cdots, y_N) $ be a $N\times 1$ vector representing one deep characteristic (a row of $\Y_{t-1}$). The weight is 
\vspace{-0.2cm}
\begin{equation}\label{eqn: weight}
\begin{aligned}
	\W_{t-1} = h^{[0]}(\y_{t-1}) &=  \underbrace{\begin{bmatrix}
		\text{softmax}(y^+_1) \\
		\text{softmax}(y^+_2) \\
		\vdots\\
		\text{softmax}(y^+_N) \\
		\end{bmatrix}}_{\text{long portfolio}} - \underbrace{\begin{bmatrix}
		\text{softmax}(y^-_1) \\
		\text{softmax}(y^-_2) \\
		\vdots\\
		\text{softmax}(y^-_N) \\
		\end{bmatrix}}_{\text{short portfolio}},\\
	\text{where } y^+ :&= -50 e^{-5 y}, \quad y^- := -50 e^{5y},\\
	\text{softmax}(y_i) &= \frac{e^{y_i}}{\sum_{i=1}^N e^{y_{i}}}.
\end{aligned}
\end{equation}
These selected tuning parameters in the definition of $y^+$ and $y^-$ guarantee that about 50\% to 70\% stocks in the middle rank have zero weight, similar to the traditional sorting procedure. Furthermore, the portfolio leverage ratio is one for the long or short leg in the deep factor. The first \text{softmax} vector in the expression of $h^{[0]}$ represents the weights of stocks in the long portfolio (large $\y_{t-1}$ leads to large weight), and the second vector represents the short portfolio (large $\y_{t-1}$ leads to small weight). To prevent the exponential operator in \text{softmax} from introducing asymmetry and exaggerating the effect of extreme values, we need to standardize $\y_{t-1}$ along the same axis and apply an additional nonlinear transformation before feeding to deep factor calculation. 

\begin{figure}[h]
	\caption{\textbf{Comparison: Weight vs. Rank}} \label{fig:weight}
	\noindent \footnotesize This figure shows the example of \text{softmax} rank weights for 3,000 stocks, $h^{[0]}(\y_{t-1}) = \text{softmax}(-50e^{-5\y_{t-1}}) - \text{softmax}(-50e^{5\y_{t-1}})$. The $x$-axis shows the cross-sectional ranks of stocks. 
	In the right panel, $y_i$'s are distributed as standard normal. The red line is the \text{softmax} weight, the blue line is the equal weight (with threshold = 1/3), and the green line is the linear rank weight. 
	In the left panel, the red line remains the same. The purple line is the \text{softmax} weights when $y_i$'s are standardized samples from $\text{LogNormal}(0,1)$. The orange line is the \text{softmax} weights when $y_i$'s are standardized samples from $\text{Uniform}(0,1)$.
	
\vspace{-0.3cm}	
	
	\begin{center}	
		\includegraphics[width=0.9\textwidth]{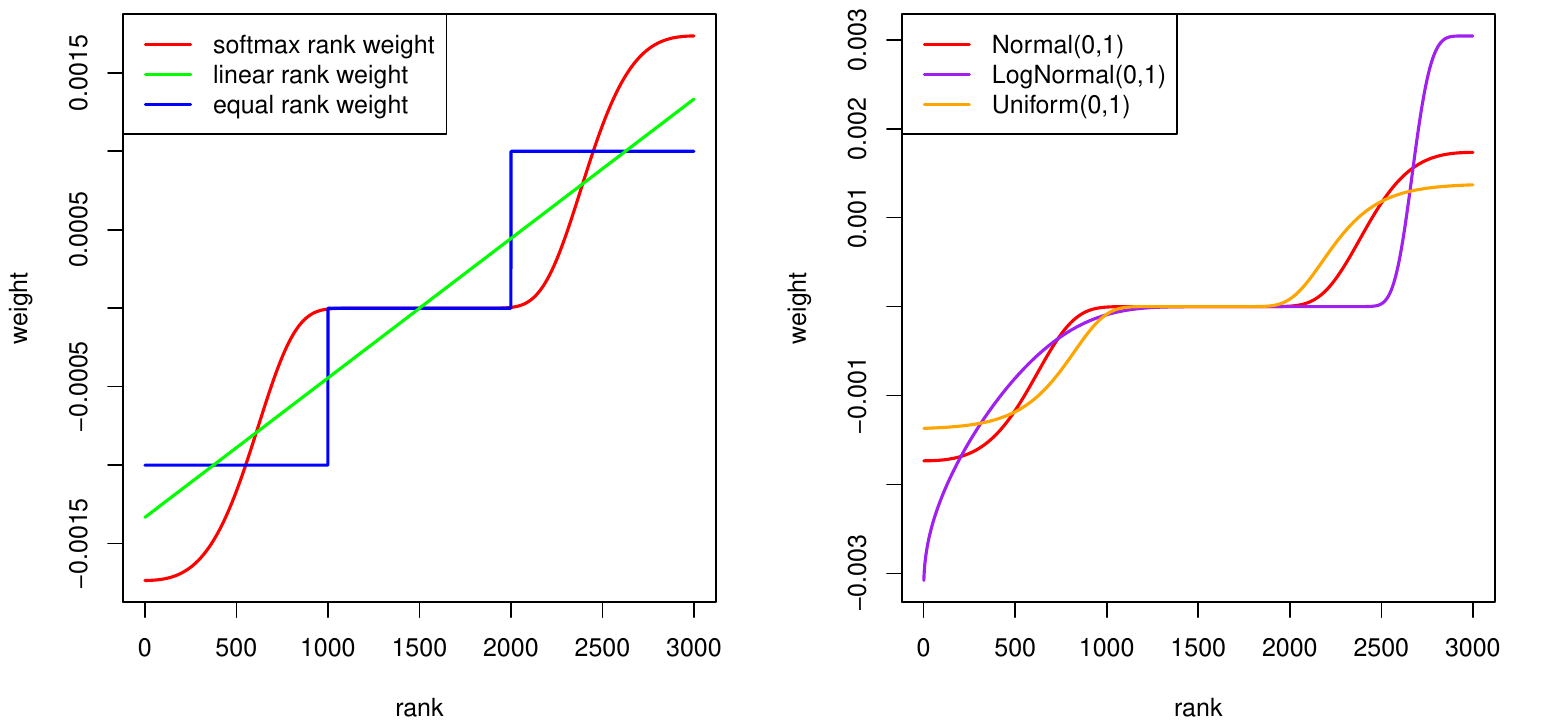}
	\end{center}
	
\vspace{-0.5cm}		
\end{figure}

Figure \ref{fig:weight} visualizes the rank-weight function with other alternatives. 
The left panel illustrates the final output of $h^{[0]}$ (red line), the portfolio weights $\W$, when $y^+ = -50e^{-5y}$, $y^- = -50e^{5y}$, and $\y=[y_1, y_2, \cdots, y_{3000}]^\intercal$ is drawn from standard normal distribution $N(0, 1)$. The choice of parameters $[-50, 5, -5]$ guarantees that the weight function has an S-shaped curve, middle-ranged assets have weight 0, and assets with the small or large rank of deep characteristics have negative and positive weights, respectively. This specific rank-weight function mimics the popular sorting strategy in empirical asset pricing.

For comparison, we also plot the standard equal weights (blue line) with the top and bottom $1/3$ of stocks as well as the rank weights introduced by \cite{frazzini2014betting} (green line). Whereas their rank weights are linear in firms' cross-sectional rankings, our weighting scheme adds nonlinearity. However, although the standard equal weights are straightforward, it is hard to incorporate them into the deep learning framework since the step function is not differentiable. 
Regarding actual holdings, \cite{novy2022betting} point out that linear rank-weighted and equal-weighted portfolios are highly overlapped (83\%). The departure of our nonlinear rank-weighted portfolio from these latter two portfolios is obvious: it ``tilts" even more toward stocks with extreme characteristics.
The flexibility of deep learning allows us to tune portfolio weights by changing the functional form of ($\y_{t-1}^+$, $\y_{t-1}^-$). 

In contrast to equal weights and linear rank weights, our \text{softmax} weights depend on the standardized distribution's cross-sectional rank information and distributional features (like skewness). 
The right panel of Figure \ref{fig:weight} presents the \text{softmax} weights when characteristics are drawn from the skewed distribution $\text{LogNormal}(1,3)$\footnote{For example, all size-related characteristics follow a lognormal distribution.} (purple line) and the bounded distribution $\text{Uniform}[0,1]$\footnote{For example, characteristics such as performance scores follow a bounded distribution.} (orange line).
Notably, the distribution of characteristics affects the symmetry and curvature of the weight curve. We see that, compared with the standard normal case, uniform characteristics lead to more holdings of stocks with middle ranks and fewer holdings of stocks at the top and bottom. 
The log-normal distribution breaks the symmetry of weights in the long and short portfolios. In this case, the long portfolio only holds a small proportion of stocks in the right tail, and the short portfolio holds almost all stocks in the lower half but still favors those in the left tail. 
Section \ref{sec: deep char} and \ref{sec: weight} jointly illustrate the deep learning transformation from lagged characteristics $\Z_{t-1}$ to individual asset weights $\W_{t-1}$. For simplicity, we refer to the transformation as 
\vspace{-0.2cm}
\begin{equation}\label{eqn: bigW} 
\W_{t-1} = H\left(\Z_{t-1}\right).
\end{equation}

\subsection{Deep Factors and Nonlinear Dynamic Betas}\label{sec: deep f}
This subsection continues with constructing deep factors based on long-short portfolio weights $\W_{t-1}$, and then an augmented deep factor model for asset pricing. We need the individual stock returns and the corresponding weights to create the long-short factors. The first $L$ layers of the model generate the deep characteristics. Then the following layers generate the portfolio weights $\W_{t-1}$ as follows,
\vspace{-0.2cm}
\begin{equation} \label{eqn: weight-DL} 
	\begin{aligned}
	    \W_{t-1} &\defeq \text{Input from Equation (\ref{eqn: weight})}\\
	    \f_{b,t} &\defeq \W_{t-1}\rr_{t}\\
	    \widehat{r}_{i,t} & \defeq \boldsymbol{\beta}\left(\z_{i, t-1}\right)^\intercal \F_t = \boldsymbol{\beta}_{d}\left(\z_{i, t-1}\right)^\intercal\f_{d,t} + \boldsymbol{\beta}_{b}\left(\z_{i, t-1}\right)^\intercal \f_{b,t}            
	\end{aligned}
\end{equation}
Given the individual stock returns $\rr_t$ and corresponding weight $\W_{t-1}$ by sorting on deep characteristics, the deep factor is simply a matrix multiplication $\f_{d,t} = \W_{t-1}\rr_t$. It defines how we construct deep factors as traded portfolios from $N$ individual stocks.

Furthermore, our deep factor $\f_{d,t}$ augments the benchmark model $\f_{b,t}$ such as CAPM or FF5, which are given exogenously as input to the augmented deep factor model. The corresponding factor loadings are $\boldsymbol{\beta}_d\left(\cdot\right)\in\mathbb{R}^{N \times P_d}$ and $\boldsymbol{\beta}_{b}\left(\cdot\right)\in\mathbb{R}^{N \times P_b}$. The factor model implied return is
\vspace{-0.2cm}
$$
\widehat{r}_{i,t} = \boldsymbol{\beta}_{d}\left(\z_{i, t-1}\right)^\intercal\f_{d,t} + \boldsymbol{\beta}_{b}\left(\z_{i, t-1}\right)^\intercal \f_{b,t}.
$$

The flexibility of deep learning allows us to model the nonlinear betas, $\boldsymbol{\beta}_{d}\left(\cdot\right)$ and $\boldsymbol{\beta}_{b}\left(\cdot\right)$, which are driven by firm characteristics $\z_{i,t -1}$ for each stock. 
The current literature on the conditional factor model mainly adopts a linear functional form for betas on firm characteristics \citep{ferson1999conditioning}, and our framework provides one of the first internal consistent nonlinear beta modelings \citep[e.g.,][]{gu2021autoencoder}.
Specifically, we assume betas follow a neural network structure\footnote{In the empirical study, we fix one three-layer neural network specification, with [64, 16, 4] neurons in each layer and \text{tanh} activation functions without dropout.} on the same set of high-dimensional characteristics for training the deep characteristic. 
\vspace{-0.2cm}
\begin{equation}
    \begin{aligned} \label{eqn: beta}
        \left[\boldsymbol{\beta}_{d}\left(\z_{i,t-1}\right)^\intercal, \boldsymbol{\beta}_{b}\left(\z_{i,t-1}\right)^\intercal\right]^\intercal &= G\left(\z_{i, t-1}\right) 
    \end{aligned}
\end{equation}
The beta neural network is jointly trained with the factor neural network to minimize the loss function. 
Without knowing portfolio characteristics, we can estimate constant betas by the time series regression in the empirical study for fitting portfolio returns.

\subsection{Minimizing Loss Function} \label{sec: loss}
Equation (\ref{eqn: bigW}) defines the deep learning transformation from lagged characteristics to portfolio weights $H\left(\cdot\right)$, which is essentially a composition function $H\left(\Z_{t-1}\right) = h^{[0]}\circ F^{[L]}\circ \cdots \circ F^{[1]}\left(\Z_{t-1}\right)$. Although the final layer of \text{softmax} activation is the key idea of interpreting the security sorting as an activation function within a deep learner, the previous layers of activation play the role of dimension reduction and variable selection under economic guidance. In the empirical study, we choose $F^{[1]} = F^{[2]} = \cdots = F^{[L]}$ to be the $\text{tanh}\left(\cdot\right)$ activation function, i.e., $F\left(x\right) = \left(e^{x} - e^{-x}\right)/\left(e^{x} + e^{-x}\right)$.\footnote{We have conducted a robustness check for switching the activation function to \text{ReLU} and found robust empirical results. The intuition is the final layer \text{softmax} activation is mainly about the rank and distribution of previous transformed inputs rather than their calculated values.}

Fixing the number of layers $L$ and the deep learning architecture $\{K_l\}_{l=1}^{L}$, our objective function is the quadratic sum of realized pricing errors regularized by absolute values of off-diagonal weights. 
\vspace{-0.2cm}
\begin{equation}\label{eqn: objective}
\mathcal{L}_\lambda\left(\widehat{\boldsymbol{\Theta}}\right) \defeq \frac{1}{NT} \sum_{t=1}^T \sum_{i=1}^N\left(r_{i,t} - \widehat{r}_{i,t}\right)^2 + \lambda \sum_{l=1}^{L-1} \sum_{i\neq j} |A^{[l]}_{i,j}|, 
\end{equation} 
where $\widehat{\boldsymbol{\Theta}}$ is the set of all parameters in the neural networks, and
\vspace{-0.2cm}
\begin{equation}\label{eqn: returns}
    \widehat r_{i,t} =  \widehat{\boldsymbol{\beta}}_d\left(\z_{i,t-1}\right)^\intercal \f_{d,t} + \widehat{\boldsymbol{\beta}}_b\left(\z_{i,t-1}\right)^\intercal \f_{b,t}
\end{equation}
$$
	\left[\widehat{\boldsymbol{\beta}}_{d}\left(\z_{i,t-1}\right)^\intercal, \widehat{\boldsymbol{\beta}}_{b}\left(\z_{i,t-1}\right)^\intercal\right]^\intercal =  G\left(\z_{i, t-1}\right),
$$

Here, $\lambda$ is the regularization parameter, which cross-validation can choose. The $L_1$ penalty of the off-diagonal weights aims to stabilize the model and keep the combination of characteristics sparse. Training the deep network is then equivalent to obtaining a joint estimation of all bias and weight terms in the neural networks generating factors $\mathbf{f}_{d,t}$ and dynamics of $\boldsymbol{\beta}(\cdot)$ by minimizing the objective function in Equation (\ref{eqn: objective}).
\vspace{-0.2cm}
$$
	\widehat{\boldsymbol{\Theta}} = {\arg\min} \,\mathcal{L}_\lambda.
$$
The quadratic sum of realized pricing errors can be decomposed into the average pricing errors (variation of unexplained average returns in the cross section) and residual errors (variation of unexplained returns in the time series). We deploy the stochastic gradient descent algorithm searching for the optimal parameters, a standard approach for deep learning. See Appendix \ref{app: optim} for optimization details.
\vspace{-0.2cm}
\begin{equation}
    \begin{aligned} \label{eqn: variation}
        \frac{1}{NT} \sum_{t=1}^T \sum_{i=1}^N\left(r_{i,t} - \widehat{r}_{i,t}\right)^2 &= \frac{1}{NT} \sum_{t=1}^T \sum_{i=1}^N\left(\widehat \alpha_{i} + \widehat \epsilon_{i,t}\right)^2 \\
        &= \underbrace{\frac{1}{N} \sum_{i=1}^N \widehat\alpha_{i}^2}_{\text{pricing errors}}  + \underbrace{\frac{1}{NT} \sum_{t=1}^T \sum_{i=1}^N \widehat{\epsilon}_{i,t}^2}_{\text{time-series variation}}
    \end{aligned}
\end{equation}

We summarize the deep learning architecture in Table \ref{tab: mechanism} and Figure \ref{fig:network}.  Empirically, we set layer size $K_l$ for the $l$-th layer equal to $K_0$ and the number of layers $1 \leq L \leq 4$. For example, a two-layer ($L=4$) network has layer sizes $K_0\rightarrow K_0 \rightarrow P_d$ from $\Z^{[0]}_{t-1}$ to $\Y_{t-1}$.

\section{Empirical Findings}\label{sec: empirical}

\subsection{Data and Implementation Design}\label{sec: emp_data}
The empirical study uses monthly data from January 1972 to December 2021. We follow the Fama-French filtering of individual stocks and use the largest 3,000 firms of lagged market equity values.\footnote{
We select stocks listed on NYSE, AMEX, or NASDAQ for over a year, observations with a CRSP share code of 10 or 11. Observations with negative book equity or lag market equity values are excluded.}
We take 60 firm characteristics from six major categories: momentum, value, investment, profitability, frictions (or size), and intangibles. We follow \cite{hou2017replicating} to calculate these characteristics but adopt many of them with quarterly or monthly available information for our monthly sorting scheme. The cross section of monthly characteristics is standardized uniformly into the $[-1,1]$ range.\footnote{For example, the cross-sectional market equity values in December 2018 are uniformly standardized into the $[-1, 1]$ range. The firm with the lowest value is -1, and the one with the highest is 1. Therefore, this uniform standardization is a monthly cross-sectional standardization that transforms the data into $[-1, 1]$. If a firm has missing values for some characteristics, the imputed values are 0, which implies the firm is not important in security sorting. The activation function on these missing values imputed by zero is still zero.}

The augmented deep factor model is trained on individual stock returns. The objective function is a conditional factor model with characteristics-driven betas. However, suppose one wants to evaluate the trained model on test portfolios. In that case, one might not know the underlying characteristics and can only consider the unconditional factor model with constant betas.
For robustness check, we consider the monthly univariate-sorted $10\times 1$ portfolios ($10 \times 1 \times 60 = 600$), monthly bivariate-sorted $3\times 5$ portfolios between size (large-, mid-, and small-cap) and other characteristics ($3 \times 5 \times 59 = 885$), Fama-French $5 \times 5$ Size-BE/ME portfolios, and Fama-French 49-industry portfolios.

To prevent overfitting, we implement four-fold cross-validation with a deterministic partition of training and validation data to determine the best tuning parameters (number of factors and regularization penalties) for each layer's augmented deep factor models.\footnote{We thank an anonymous referee for suggesting a longer cross-validation period. The previous version of the paper considers a validation scheme using a different set of test assets.}
We consider two deterministic validation samples for reserving the complete cross-sectional dependence structure: 1972-1991 and 1992-2011.
The consecutive periods in the validation sample also help reserve factor and beta explanatory power continuity.
The recent decade from 2012 to 2021 is used for the test window.\footnote{
Because of the larger noise in individual stock returns, we consider winsorization when working with individual stock returns to reduce the impact of outliers. We perform monthly 2.5\% and 97.5\% cross-sectional winsorization on returns in the training sample from 1972 to 2011.}

\subsection{Performance Measures}\label{sec: measure}
We follow \cite{kelly2019characteristics} to include multiple performance measures for economic and statistical purposes. Total $R^2$ and Predictive $R^2$ are designed for measuring statistical model fitness, which can be used for individual stocks and test portfolios. 
Cross-sectional $R^2$ are designed for evaluating economic asset pricing performance, the non-arbitrage condition for zero alphas, which can only be used for test portfolios. 
\cite{kelly2019characteristics} compare all these model performance measures with the zero benchmark. 
If one uses the excess market or CAPM-implied return for calculating the denominator, a positive $R^2$ indicates the outperformance over the stronger market-based benchmarks.\footnote{We thank an anonymous referee for suggesting the market-based benchmarks.}

\paragraph{Performance Measures for Individual Stocks}
\vspace{-0.5cm}
\begin{equation}\label{eqn: totalR2_ind}
	\text{Total }R^2 = 1 - \frac{\sum_{i=1}^{N}\sum_{t=1}^{T} \left(r_{i,t} - \widehat{r}_{i,t}\right)^2}{\sum_{i=1}^{N}\sum_{t=1}^{T} \left(r_{i,t} - \text{MktRF}_t\right)^2},	
\end{equation}
where $\widehat{r}_{i,t} = \widehat{\boldsymbol{\beta}}_{d}\left(\z_{i,t-1}\right)^\intercal \f_{d,t} + \widehat{\boldsymbol{\beta}}_{b}\left(\z_{i,t-1}\right)^\intercal \f_{b,t}$. Total $R^2$ represents the fraction of realized return variation explained by the factor model-implied contemporaneous return, aggregated over all assets and all periods. 
Our objective function directly relates to this Total $R^2$ for minimizing aggregated realized pricing errors.
	
\vspace{-0.5cm}
\begin{equation}\label{eqn: predR2_ind}
	\text{Predictive }R^2 = 1 - \frac{ \sum_{i=1}^{N}\sum_{t=1}^{T_i} \left(r_{i,t} - \widehat{r}_{i,t}\right)^2}{\sum_{i=1}^{N}\sum_{t=1}^{T_i} \left(r_{i,t} - \lambda_{\text{MktRF}}\right)^2},	
\end{equation}
where $\widehat{r}_{i,t} = \widehat{\boldsymbol{\beta}}_{d}\left(\z_{i,t-1}\right)^\intercal \boldsymbol{\lambda}_{f_d} + \widehat{\boldsymbol{\beta}}_{b}\left(\z_{i,t-1}\right)^\intercal \boldsymbol{\lambda}_{f_b}$. The risk premia $\boldsymbol{\lambda}_{f_d}$ and $\boldsymbol{\lambda}_{f_b}$ are the risk premia estimate, which can be the average returns of traded factors. Predictive $R^2$ summarizes the predictive performance by the factor model-implied return forecasts, aggregated over all assets and all periods.

\paragraph{Performance Measures for Test Portfolios}
One might not know the underlying characteristics of test portfolios without knowing portfolio weights. Thus, we estimate an unconditional factor model with constant betas $\widehat{\boldsymbol{\beta}}_{d,i}$ and $\widehat{\boldsymbol{\beta}}_{b,i}$ by the standard time series regressions. 
\vspace{-0.2cm}
	      \begin{equation}\label{eqn: totalR2}
	      	\text{Total }R^2 = 1 - \frac{\sum_{t=1}^T \sum_{i=1}^N\left(r_{i,t} - \widehat{r}_{i,t}\right)^2}{\sum_{t=1}^T \sum_{i=1}^N \left(r_{i,t} - \widehat{R}^{CAPM}_{i,t}\right)^2},	
	      \end{equation}
	      where $\widehat{r}_{i,t} = \widehat{\boldsymbol{\beta}}_{d,i}^\intercal \f_{d,t} + \widehat{\boldsymbol{\beta}}_{b,i}^\intercal \f_{b,t}$ and $\widehat{R}^{CAPM}_{i,t}$ is the CAPM implied return.

\vspace{-0.5cm}	
	      \begin{equation}\label{eqn: predR2}
	      	\text{Predictive }R^2 = 1 - \frac{\sum_{t=1}^T \sum_{i=1}^N\left(r_{i,t} - \widehat{r}_{i,t}\right)^2}{\sum_{t=1}^T \sum_{i=1}^N \left(r_{i,t}  - \widehat{ER}^{CAPM}_{i}\right)^2},	
	      \end{equation}
	      where $\widehat{r}_{i,t} = \widehat{\boldsymbol{\beta}}_{d,i}^\intercal \widehat{\boldsymbol{\lambda}}_{f_d} + \widehat{\boldsymbol{\beta}}_{b,i}^\intercal \widehat{\boldsymbol{\lambda}}_{f_b}$ and $\widehat{ER}^{CAPM}_{i}$ is the CAPM implied average return with the constant market risk premium $ \boldsymbol{\widehat \lambda}_{\text{MktRF}}$. The risk premium estimation adopts the time series average of risk factor returns.

\vspace{-0.5cm}	      
	      \begin{equation}\label{eqn: csR2}
	      	\text{Cross-sectional }R^2 = 1 - \frac{\sum_{i=1}^N\left( \bar r_{i} - \widehat{\bar r_{i}}\right)^2}{\sum_{i=1}^N \left(\bar r_{i}   - \widetilde{ER}^{CAPM}_{i} \right)^2},	
	      \end{equation}
	      where $\widehat{\bar r_{i}} = \widehat{\boldsymbol{\beta}}_{d,i}^\intercal \widetilde{\boldsymbol{\lambda}}_{f_d} + \widehat{\boldsymbol{\beta}}_{b,i}^\intercal \widetilde{\boldsymbol{\lambda}}_{f_b}$ and $\widetilde{R}^{CAPM}_{i,t}$ is the CAPM implied average return with the constant market risk premium $\boldsymbol{\widetilde \lambda}_{\text{MktRF}}$. The risk premium estimation adopts the cross-sectional regression estimates of risk factors.
	      The cross-sectional $R^2$ represents the faction of assets' average returns explained by the model-implied expected returns.

\subsection{Asset Pricing and Statistical Performance} \label{sec: ap}
\paragraph{Pricing Individual Stocks}
Columns labeled by ``12-21" in Table \ref{tab: R2_ind} summarize the pricing performance of individual stock returns, which evaluates the model on the period ``12-21" and trains the model using forty years of the other four periods. 
We construct deep factors to augment two benchmark models, CAPM and FF5. 
We perform the four-fold deterministic cross-validation to determine the optimal number of factors.\footnote{The four folds are deterministic for four consecutive decades over time, rather than random sampling. 
In addition to deep learning tuning parameters, we try different numbers of factors and layers for the model selection. 
The highlighted two-layer model is selected for the CAPM benchmark, and the highlighted three-layer model is selected for FF5. Both include five additional deep factors that augment the benchmarks.} 
We further compare with two strong models, Risk-Premium Principal Component Analysis (RP-PCA) of \cite{lettau2018estimating}, and Instrumental Principal Component Analysis (IPCA) of \cite{kelly2019characteristics}, both include high-dimensional characteristics in estimating principal components using the same sample.\footnote{Five RP-PCA factors are estimated using 600 10$\times$1 univariate-sorted basis portfolios, and five IPCA factors are generated on individual stock returns using 60 rank characteristics.} 


We use the excess market return as the benchmark for Total $R^2$ in Equation (\ref{eqn: totalR2_ind}). A positive value indicates a better pricing performance than the market. 
First, adding deep factors leads to a substantial increase in performance, which is consistent for models with different layers on either CAPM or FF5 benchmarks. 
Second, IPCA and RP-PCA are strong benchmark models and consistently outperform CAPM and FF5 for both in-sample and out-of-sample periods. 
All augmented deep factor models for different layers show excellent out-of-sample performance using in-sample estimated beta functions. 
For example, our selected three-layer model on FF5, can achieve as high as 8.16\% in sample Total $R^2$ and 5.51\% out-of-sample, both of which are second highest. 
However, the weaker results for augmented deep factor models on CAPM (one benchmark factor) indicate the limitations of deep learning given the short history of financial data.

The excess market return is a much stronger benchmark than zero for Predictive $R^2$ in Equation (\ref{eqn: predR2_ind}),\footnote{\cite{gu2020empirical} use the zero benchmark for calculating their out-of-sample $R^2$.} which is similar to out-of-sample $R^2$ for evaluating return predictability but requires a factor model structure. 
The economic-guided factor model structure is a strong constraint by forcing the predictability from risk factors. 
All augmented deep factor models on CAPM and FF5 consistently provide positive Predictive $R^2$ values for both in-sample and out-of-sample periods.
The selected three-layer model on FF5 makes 0.48\% and 0.25\%, respectively. 
Its out-of-sample performance for Predictive $R^2$ is better than FF5 and IPCA, though slightly lower than the RP-PCA.

\paragraph{Pricing Test Portfolios} 

Tables \ref{tab: portR2_ins} and \ref{tab: portR2_oos} report Total, Predictive, and Cross-sectional $R^2$ for four sets of test portfolios: 600 univariate-sorted, 885 bivariate-sorted, 25 Size-BE/ME, and 49 industries, which have varying signal-to-noise ratios for return variations and help demonstrate the model performance robustness. 
We use the CAPM implied or average returns as the benchmark for calculating these values. Therefore, a positive value indicates a better pricing performance than CAPM. 
Note that we can only refit an unconditional factor model with constant betas for portfolios with unknown weights.
The portfolio betas are estimated by time-series regression and fixed for the out-of-sample analysis.


Our results show that the FF5 model mainly outperforms CAPM for three sets of characteristics-sorted portfolios on the Total $R^2$. 
In particular, adding deep factors leads to consistent performance improvement on benchmark models with different layers of deep learning structures on CAPM or FF5 benchmarks. 
All other methods in Panel C have weak performance on some test portfolios. Still, our augmented deep factor models on FF5 consistently price all types of test portfolios for both in-sample and out-of-sample periods.
These findings are consistent with those of \cite*{lewellen2010skeptical} that long-short factors help to explain characteristics-sorted portfolios, which tend to have a stronger factor structure. 
Our augmented deep factor models still reasonably price 49 industry portfolios, but PCA methods have mixed performance. 
The patterns of industry portfolios may not be easy to learn from pure latent factor generation on characteristics, so including the market factor benchmark might be important to the latent factor generation.


We also provide the Cross-sectional $R^2$ in Tables \ref{tab: portR2_ins} and \ref{tab: portR2_oos}, which explains the cross-sectional variation in average returns by the model-implied expected returns. 
If the model is sufficiently well, the model-implied expected return should align with the average returns.  
Our augmented deep factor models on either CAPM or FF5 provide consistently positive performance to all types of test portfolios for both in-sample and out-of-sample analysis, with most numbers higher than those of FF5. 
For Predictive $R^2$, the return forecasting performance of FF5 and CAPM is close. 
The mixed and weak results in Predictive $R^2$ for all these factor models suggest that predicting portfolio returns is difficult when restricting the factor structure. 
Finally, the relatively weaker but still mostly positive results for augmented deep factor models on CAPM indicate the limitations of deep learning given the short history of financial data. 
We are combining deep learning and human learning by including those known useful factors to achieve better model fitness.

\subsection{Investing Deep Factors}\label{sec: invest-deep}
Asset pricing theory highlights the equivalence between the mean-variance efficient (MVE) portfolio and the stochastic discount factor (SDF), where the minimum variance of the SDF corresponds to the squared maximal Sharpe ratio of the MVE portfolio \citep{HJ1991JPE}. 
As such, explaining the cross section and maximizing the Sharpe ratio are direct and indirect objectives of the SDF. 
This subsection presents how to use the generated deep factors and build a factor investing portfolio in improving benchmark factors' performance. 
\cite{kozak2020shrinking} also show the portfolio performance for SDF coefficients on factors, which is equivalent to the mean-variance efficient or tangency portfolio weights: 
	$b = \Sigma_F^{-1} \mu_f$, 
where $\f_t = \left[\f_{d,t}^\intercal, \f_{b,t}^\intercal\right]^\intercal$. The MVE portfolio is $b^\intercal\f_t$.

Table \ref{tab: investment} presents the annualized Sharpe ratios for factor investing portfolios constructed using the MVE portfolio weights $b$ estimated from the training sample of 40 years, which are then fixed for out-of-sample evaluation. 
Though the direct objective is asset pricing performance, the in-sample risk-adjusted Sharpe ratios are mostly above 3 for all augmented deep factor models on the training sample from 1972 to 2011. 
We include the annualized Sharpe ratios of the CAPM and Fama-French 5-factor models for comparison.
Our results show highly positive and consistent Sharpe ratios for RP-PCA and IPCA in the 40-year training sample. IPCA has produced the highest Sharpe ratio of 3.67 in the recent decade, while our 2-layer DL-CAPM and 3-layer DL-FF5 augmented model achieves comparable Sharpe ratios of 2.49 and 3.00. 
Our deep factors are generated to complement benchmark models and consistently improve in-sample and out-of-sample investment performance across different layers of architectures.


Table \ref{tab: spanreg} provides the spanning regression results for deep factors generated on the CAPM or FF5 benchmarks corresponding to the selected factor models.
The analysis uses all 50 years of data, while the deep factors are only generated with the first forty years of data.
First of all, the MVE portfolio for CAPM (2-layer) and FF5 (3-layer) benchmarks Table \ref{tab: investment} have extremely high model-adjusted alphas (above 4\% monthly) on Fama-French five factors. 
The time series explained $R^2$ are low in both panels, which suggests the MVE portfolio of constructed deep factors offers perfect hedging performance on the benchmark models. 
The benchmark hedging conclusion is consistent with the deep tangency portfolio in \cite*{feng2022DTP}.
Second, we further decompose each deep factor and evaluate their loadings on Fama-French five factors.
Deep learning generates these deep factors in a batch, and there is no order for these factors. 
Because we restrict the long and short portfolio weights to one, it is unsurprising to find moderate beta values. 


\subsection{Robustness for Rotating Test Samples} \label{sec: robustness}
Our main analysis evaluates our model performance using data from the recent decade as the test sample. To ensure our findings' robustness, we rotate train and test samples across the entire dataset, following \cite{kozak2020shrinking} and \cite*{kaniel2022machine}.
We divide the data into five consecutive folds and designate each decade of data as the test sample, and implement a deterministic four-fold cross-validation using the rest four decades of data for training the model.
This approach allows us to evaluate the model performance under different macroeconomic conditions each decade.

In Table \ref{tab: R2_ind}, though the ``out-of-sample" model explanatory power varies from different decades, results for all these factor models are consistently positive for all ``in-sample" model fitting. 
This implies that the factor structures of cross-sectional individual stock returns are relatively stable, demonstrating the robustness of our augmented deep factor model. 
Using future data to predict past data in 72-81 provides the best numbers for Total and Predictive $R^2$ to all models. 
The augmented deep factor models on FF5 provide the second-best results next to IPCA.

We also assess the robustness of investment performances in Table \ref{tab: investment}. Though in-sample results for Fama-French five factors are consistent and slightly above one, the results for the MVE portfolio of deep factors plus the benchmark factors are robustly high, mostly from 3 to 5,  for each subsample. 
The out-of-sample investment performances are also robust. For example, for two selected factor models on CAPM and FF5, numbers in different decades are similar to the positive ones in the recent decade.

\subsection{Interpreting Deep Characteristics}\label{sec: interpret-char}
\begin{figure}
	\caption{Deep Characteristics v.s. Fama-French Characteristics} \label{fig: charplot}
	\noindent \footnotesize This figure provides the fitted smooth splines for our ranked deep characteristics (augmented 2-layer 5-factor model on CAPM) to four Fama-French ranked characteristics. These smooth splines are plotted at the firm level in the cross section, and dotted lines are the corresponding 95\% confidence intervals. 
	\begin{center}
		\includegraphics[width=\linewidth]{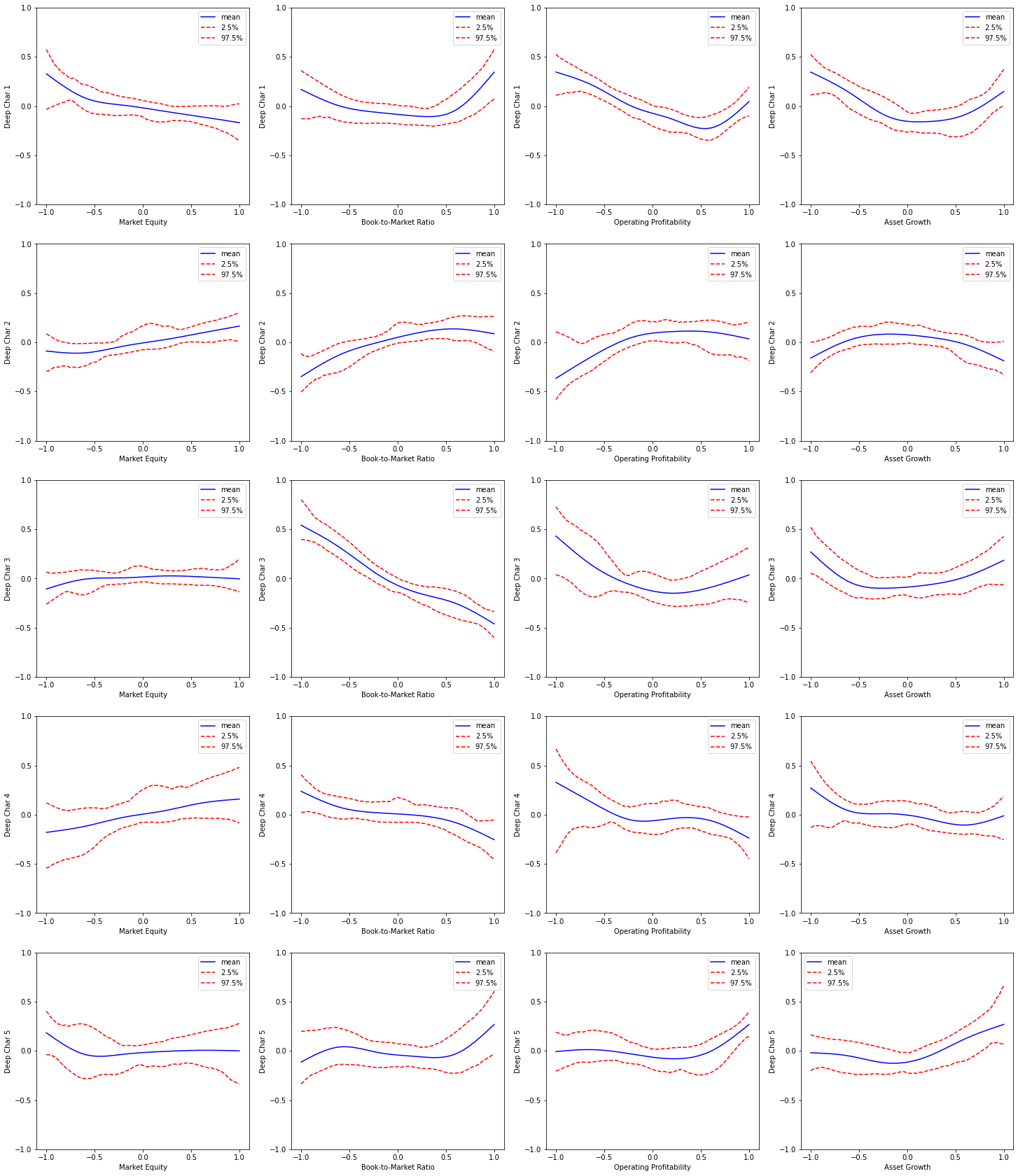}
	\end{center}
\end{figure}

We examine the economic interpretation of deep characteristics that generate deep factors, which can be utilized for lower frequency sorting of securities, reducing transaction costs. 
Our structural mechanism offers an advantage over factor models, allowing visualization of the nonlinear relationships among characteristics and understanding how raw characteristics drive latent risk factors. 
The \text{tanh}$(\cdot)$ activation function directly transforms raw characteristics into deep characteristics in the neural network. 
Figure \ref{fig: charplot} shows unique empirical outcomes through interaction and nonlinear transformation of characteristics, while the literature typically focuses on factors.


We examine a 2-layer augmented deep factor model on CAPM and five additional deep factors to gain insights into firm characteristics. Using smooth splines and 95\% confidence intervals, we plot the fitted data for the Fama-French four characteristics: market equity, book-to-market ratio, operating profitability, and investment. 
One of the deep factors shows an exceptionally high alpha value that the Fama-French five factors cannot explain. We investigate its underlying characteristics and find it asymmetrically exposed to small-cap, high-profitability, and high-investment stocks. However, it has symmetric exposures to both value and growth stocks. 
The first deep characteristic demonstrates U-shaped relationships with book-to-market ratios, operating profitability, and asset growth, while the second characteristic displays an inverse U-shaped relationship with these same factors. 
Notably, our characteristics data are uniformly distributed from -1 to 1, and these plots demonstrate the nonlinear patterns in our augmented deep factor model, providing valuable insights into asset pricing.
    
\paragraph{Nonlinear Exposures}  
While the constructed deep characteristics differ from ordered principal components, we can still evaluate the contribution of each raw rank characteristic to their construction, as well as the deep factor loadings. 
Deep learning optimization allows one to evaluate the importance of variables by calculating the sensitivity of the final loss function or one of the hidden layers to the input variables. 
To understand the sources of deep characteristics and factor betas, which are hidden layers in our deep learning architecture, we calculate their nonlinear exposures from 60 raw characteristics.
\cite{chen2019deep} also use a similar measure to calculate the value from the loss function to characteristics, while we use the aggregated absolute gradient value from the deep characteristic and betas to raw characteristics 

\vspace{-0.5cm}	
\begin{equation}\label{eqn: grad_char}
\text{Importance}_{\mathbf{Y_p}}(z_j) = \frac{1}{NT} \sum_{i=1}^N \sum_{t=1}^T \left| \frac{\partial Y_{i,p,t}}{\partial z_{i,j,t}}\right|
\end{equation}

\vspace{-0.5cm}
\begin{equation} \label{eqn: grad_beta}
\text{Importance}_{\beta}(z_j) =\frac{1}{NT(P_d + P_b)} \sum_{i=1}^N\sum_{t=1}^T \sum_{p=1}^{(P_d + P_b)} \left|\frac{\partial \beta_{i,p,t}}{\partial z_{i,j,t-1}}\right|,
\end{equation}
where $Y_{i,p,t}$ is the $p$-th deep characteristic generated for asset $i$ at period $t$, and $\beta_{i,p,t}$ is the $p$-th factor loading for asset $i$ at period $t$. Notice that, as described in Section \ref{sec: deep char}, the activation function for the deep characteristic generation is univariately operated, so characteristic values of individual assets do not affect others.

\begin{figure}
\caption{Nonlinear Exposures for Deep Characteristics and Betas} \label{fig: gradient}
	\noindent \footnotesize This figure provides the nonlinear exposures to the deep characteristics and factor betas construction from 60 raw characteristics. 
    We calculate the absolute gradient values (see Equation \ref{eqn: grad_char} and \ref{eqn: grad_beta}) for the fifth deep characteristic on the CAPM benchmark and the aggregated characteristics-driven nonlinear betas. 
    \begin{center}
    \includegraphics[width=\linewidth]{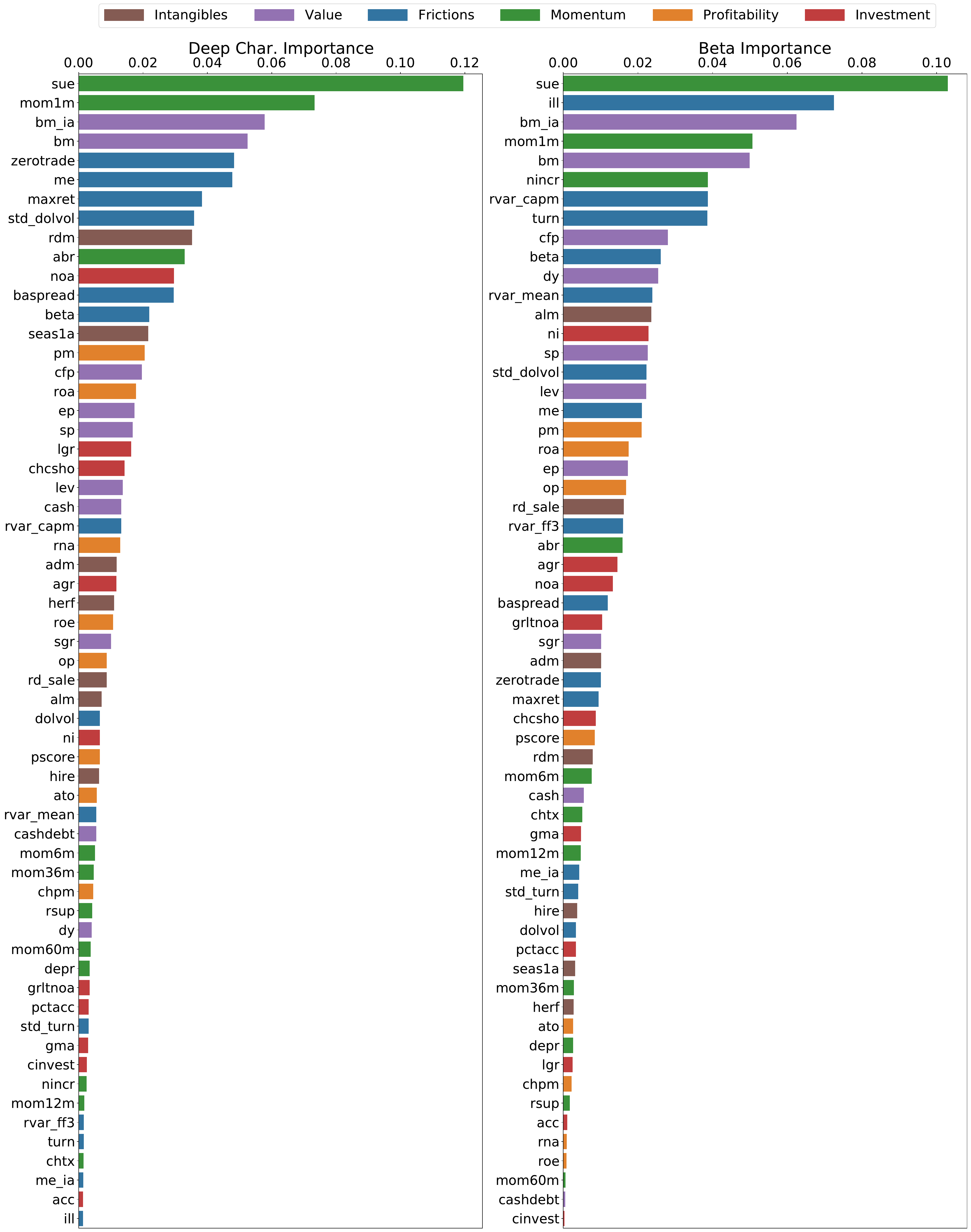}
    \end{center}
\end{figure}

Figure \ref{fig: gradient} provides the absolute gradient values for the fifth deep characteristic on the CAPM benchmark and the aggregated factor loadings.
On the one hand, several important types of characteristics, such as momentum (sue and \emph{mom1m}), value (\emph{bm\_ia} and \emph{bm}), and frictions (\emph{zerotrade} and \emph{me}).
In addition to the momentum characteristics, the nonlinear activation of book-to-market ratios and market equity values still provides incremental information.
On the other hand, the most commonly used characteristic groups for conditional factor loadings of individual stocks are momentum (sue and \emph{mom1m}), frictions (\emph{ill}), and value (\emph{bm\_ia} and \emph{bm}).
It is unsurprising to find the same group of characteristics useful for constructing characteristics and factor loadings \citep{kelly2019characteristics}.


\paragraph{Linear Exposures} 
We run the firm-level Fama-Macbeth cross-sectional regressions with deep characteristics (augmented 2-layer 5-factor model on CAPM) on all raw characteristics for the linear exposures to the deep characteristics. With 600 months of data, we calculate the average regression slope and provide coefficients and significance in Table \ref{tab: exposures_char} to demonstrate the variable importance.

\vspace{-1cm}	
\begin{equation}
	\text{deepchar}_{i,t} = a_t + b_{1,t}\text{rawchar}_{1,i,t} + \cdots + b_{60,t}\text{rawchar}_{60,i,t} + \epsilon_{i,t}
\end{equation}
We perform the cross-sectional regression for every month $t$ and calculate the Newey-West standard error with 12 lags to determine the average significance.

Table \ref{tab: exposures_char} presents the linear exposures to the deep characteristics obtained from 60 raw characteristics. The table is sorted based on the exposure values to the fifth deep characteristic, and the top five crucial raw characteristics, namely sue, \emph{mom1m}, \emph{bm\_ia}, \emph{zerotrade}, and \emph{bm}, are highlighted. These raw characteristics are highly consistent with the important characteristics for nonlinear exposures illustrated in Figure \ref{fig: charplot}, providing indirect evidence of the effectiveness of our nonlinear and linear exposure calculations. 
Moreover, the sharp decrease in magnitudes of linear exposure values aligns with the nonlinear exposure values, but a large number of significant raw characteristics may indicate unclear evidence of sparsity.


Our research yields new empirical findings by utilizing dimension reduction techniques on characteristics, as opposed to factors. While PCA focuses on maximizing variation in the cross-sectional returns, our approach directly relates raw characteristics to the objective function. Through our augmented deep factor model, we can achieve superior results compared to traditional security sorting methods. 
However, concluding whether these characteristic signals are sparse based on linear or nonlinear exposure demonstrations is challenging, as demonstrated in \cite{kozak2020shrinking}. 
Nonetheless, our deep learning transformation is valuable for the characteristics-sorted factor model.

\section{Summary}\label{sec: summary}
This paper offers an alternative perspective by constructing a structural deep neural network in empirical asset pricing. 
We adopt the deep learning framework with a bottom-up approach, which provides a complete mechanism for the characteristics-sorted factor model. By minimizing aggregated realized pricing errors, we train an augmented deep factor model using firm characteristics \emph{[inputs]}, and generate risk factors \emph{[intermediate features]} to fit the cross section of individual stock returns \emph{[outputs]}.

This paper is the first to recognize the characteristics-sorted factor model as a deep neural network and provides a systematic approach for the implementation.
Our paper is not directly related to the literature on predicting asset returns using machine learning. 
The current prediction literature studies the predictive performance between firm characteristics and security returns, but often skips the intermediate channel involved with risk factors. Our bottom-up approach fills in this missing piece.

Technically, we customize the \text{softmax} activation to approximate the long-short portfolio weights for individual assets. 
Sorting firm characteristics to generate long-short factors is underappreciated nonlinear modeling for linear factor models because researchers usually assume these are observable. 
Our method's dimension reductions are based on characteristics, which differ from the current literature that reduces the dimension of factors or basis portfolios.
Recently, \cite{feng2022DTP} consider a similar framework for constructing the tangency portfolio using individual assets.

\vspace{1cm}
{\onehalfspacing
	\bibliography{DL-alpha}

\begin{thebibliography}{}

\bibitem[\protect\citeauthoryear{Bali, Huang, Jiang, and Wen}{Bali
  et~al.}{2021}]{bali2021ml}
Bali, T.~G., D.~Huang, F.~Jiang, and Q.~Wen (2021).
\newblock {Different Strokes: Return Predictability Across Stocks and Bonds
  with Machine Learning and Big Data}.
\newblock Technical report, Georgetown University.

\bibitem[\protect\citeauthoryear{Bianchi, B{\"u}chner, and Tamoni}{Bianchi
  et~al.}{2021}]{bianchi2021bond}
Bianchi, D., M.~B{\"u}chner, and A.~Tamoni (2021).
\newblock Bond risk premiums with machine learning.
\newblock {\em Review of Financial Studies\/}~{\em 34\/}(2), 1046--1089.

\bibitem[\protect\citeauthoryear{Bryzgalova, Huang, and Julliard}{Bryzgalova
  et~al.}{2023}]{bryzgalova2021bayesian}
Bryzgalova, S., J.~Huang, and C.~Julliard (2023).
\newblock Bayesian solutions for the factor zoo: We just ran two quadrillion
  models.
\newblock {\em The Journal of Finance\/}~{\em 78\/}(1), 487--557.

\bibitem[\protect\citeauthoryear{Bryzgalova, Pelger, and Zhu}{Bryzgalova
  et~al.}{2020}]{bryzgalova2020forest}
Bryzgalova, S., M.~Pelger, and J.~Zhu (2020).
\newblock {Forest through the trees: {B}uilding cross-sections of stock
  returns}.
\newblock Technical report, London Business School.

\bibitem[\protect\citeauthoryear{Carhart}{Carhart}{1997}]{carhart1997persistence}
Carhart, M.~M. (1997).
\newblock On persistence in mutual fund performance.
\newblock {\em Journal of Finance\/}~{\em 52\/}(1), 57--82.

\bibitem[\protect\citeauthoryear{Chen, Pelger, and Zhu}{Chen
  et~al.}{2022}]{chen2019deep}
Chen, L., M.~Pelger, and J.~Zhu (2022).
\newblock Deep learning in asset pricing.
\newblock {\em Management Science, Forthcoming\/}.

\bibitem[\protect\citeauthoryear{Chinco, Neuhierl, and Weber}{Chinco
  et~al.}{2021}]{chinco2021estimating}
Chinco, A., A.~Neuhierl, and M.~Weber (2021).
\newblock Estimating the anomaly base rate.
\newblock {\em Journal of Financial Economics\/}~{\em 140\/}(1), 101--126.

\bibitem[\protect\citeauthoryear{Cochrane}{Cochrane}{2011}]{cochrane2011presidential}
Cochrane, J.~H. (2011).
\newblock {Presidential address: Discount rates}.
\newblock {\em Journal of Finance\/}~{\em 66\/}(4), 1047--1108.

\bibitem[\protect\citeauthoryear{Cong, Feng, He, and He}{Cong
  et~al.}{2022}]{cong2021asset}
Cong, L.~W., G.~Feng, J.~He, and X.~He (2022).
\newblock Asset pricing with panel tree under global split criteria.
\newblock Technical report, City University of Hong Kong.

\bibitem[\protect\citeauthoryear{Cong, Feng, He, and Li}{Cong
  et~al.}{2022}]{cong2022uncommon}
Cong, L.~W., G.~Feng, J.~He, and J.~Li (2022).
\newblock {Uncommon Factors for Bayesian Asset Clusters}.
\newblock Technical report, City University of Hong Kong.

\bibitem[\protect\citeauthoryear{Cong, Tang, Wang, and Zhang}{Cong
  et~al.}{2020}]{cong2020alphaportfolio}
Cong, L.~W., K.~Tang, J.~Wang, and Y.~Zhang (2020).
\newblock Alpha{P}ortfolio: {D}irect {C}onstruction {T}rough {D}eep
  {R}einforcement {L}earning and {I}nterpretable {AI}.
\newblock Technical report, Cornell University.

\bibitem[\protect\citeauthoryear{De~Bondt and Thaler}{De~Bondt and
  Thaler}{1985}]{de1985does}
De~Bondt, W.~F. and R.~Thaler (1985).
\newblock Does the stock market overreact?
\newblock {\em Journal of Finance\/}~{\em 40\/}(3), 793--805.

\bibitem[\protect\citeauthoryear{DeMiguel, Martin-Utrera, Nogales, and
  Uppal}{DeMiguel et~al.}{2020}]{demiguel2020transaction}
DeMiguel, V., A.~Martin-Utrera, F.~J. Nogales, and R.~Uppal (2020).
\newblock A transaction-cost perspective on the multitude of firm
  characteristics.
\newblock {\em Review of Financial Studies\/}~{\em 33\/}(5), 2180--2222.

\bibitem[\protect\citeauthoryear{Dong, Li, Rapach, and Zhou}{Dong
  et~al.}{2022}]{dong2022anomalies}
Dong, X., Y.~Li, D.~E. Rapach, and G.~Zhou (2022).
\newblock Anomalies and the expected market return.
\newblock {\em Journal of Finance\/}~{\em 77\/}(1), 639--681.

\bibitem[\protect\citeauthoryear{Fama and French}{Fama and
  French}{1993}]{fama1993common}
Fama, E.~F. and K.~R. French (1993).
\newblock Common risk factors in the returns on stocks and bonds.
\newblock {\em Journal of Financial Economics\/}~{\em 33\/}(1), 3--56.

\bibitem[\protect\citeauthoryear{Fama and French}{Fama and
  French}{2015}]{FAMA2015}
Fama, E.~F. and K.~R. French (2015).
\newblock A five-factor asset pricing model.
\newblock {\em Journal of Financial Economics\/}~{\em 116\/}(1), 1 -- 22.

\bibitem[\protect\citeauthoryear{Fan, Ke, Liao, and Neuhierl}{Fan
  et~al.}{2022}]{fan2022structural}
Fan, J., Z.~T. Ke, Y.~Liao, and A.~Neuhierl (2022).
\newblock Structural deep learning in conditional asset pricing.
\newblock Technical report.

\bibitem[\protect\citeauthoryear{Fan, Feng, Fulop, and Li}{Fan
  et~al.}{2022}]{fan2022real}
Fan, Y., G.~Feng, A.~Fulop, and J.~Li (2022).
\newblock {Real-Time Macro Information and Bond Return Predictability: A
  Weighted Group Deep Learning Approach}.
\newblock Technical report, City University of Hong Kong.

\bibitem[\protect\citeauthoryear{Feng, Giglio, and Xiu}{Feng
  et~al.}{2020}]{feng2020taming}
Feng, G., S.~Giglio, and D.~Xiu (2020).
\newblock Taming the factor zoo: {A} test of new factors.
\newblock {\em Journal of Finance\/}~{\em 75\/}(3), 1327--1370.

\bibitem[\protect\citeauthoryear{Feng and He}{Feng and
  He}{2022}]{feng2022factor}
Feng, G. and J.~He (2022).
\newblock Factor investing: {A} {B}ayesian hierarchical approach.
\newblock {\em Journal of Econometrics\/}~{\em 230\/}(1), 183--200.

\bibitem[\protect\citeauthoryear{Feng, Jiang, Li, and Song}{Feng
  et~al.}{2022}]{feng2022DTP}
Feng, G., L.~Jiang, J.~Li, and Y.~Song (2022).
\newblock {Deep Tangency Portfolios}.
\newblock Technical report, City University of Hong Kong.

\bibitem[\protect\citeauthoryear{Ferson and Harvey}{Ferson and
  Harvey}{1999}]{ferson1999conditioning}
Ferson, W.~E. and C.~R. Harvey (1999).
\newblock Conditioning variables and the cross section of stock returns.
\newblock {\em Journal of Finance\/}~{\em 54\/}(4), 1325--1360.

\bibitem[\protect\citeauthoryear{Frazzini and Pedersen}{Frazzini and
  Pedersen}{2014}]{frazzini2014betting}
Frazzini, A. and L.~H. Pedersen (2014).
\newblock Betting against beta.
\newblock {\em Journal of Financial Economics\/}~{\em 111\/}(1), 1--25.

\bibitem[\protect\citeauthoryear{Freyberger, Neuhierl, and Weber}{Freyberger
  et~al.}{2020}]{freyberger2020dissecting}
Freyberger, J., A.~Neuhierl, and M.~Weber (2020).
\newblock Dissecting characteristics nonparametrically.
\newblock {\em Review of Financial Studies\/}~{\em 33\/}(5), 2326--2377.

\bibitem[\protect\citeauthoryear{Gibbons, Ross, and Shanken}{Gibbons
  et~al.}{1989}]{gibbons1989test}
Gibbons, M.~R., S.~A. Ross, and J.~Shanken (1989).
\newblock A test of the efficiency of a given portfolio.
\newblock {\em Econometrica\/}, 1121--1152.

\bibitem[\protect\citeauthoryear{Green, Hand, and Zhang}{Green
  et~al.}{2017}]{green2017characteristics}
Green, J., J.~R. Hand, and X.~F. Zhang (2017).
\newblock The characteristics that provide independent information about
  average us monthly stock returns.
\newblock {\em Review of Financial Studies\/}~{\em 30\/}(12), 4389--4436.

\bibitem[\protect\citeauthoryear{Gu, Kelly, and Xiu}{Gu
  et~al.}{2020}]{gu2020empirical}
Gu, S., B.~Kelly, and D.~Xiu (2020).
\newblock Empirical asset pricing via machine learning.
\newblock {\em Review of Financial Studies\/}~{\em 33\/}(5), 2223--2273.

\bibitem[\protect\citeauthoryear{Gu, Kelly, and Xiu}{Gu
  et~al.}{2021}]{gu2021autoencoder}
Gu, S., B.~Kelly, and D.~Xiu (2021).
\newblock Autoencoder asset pricing models.
\newblock {\em Journal of Econometrics\/}~{\em 222\/}(1), 429--450.

\bibitem[\protect\citeauthoryear{Han, He, Rapach, and Zhou}{Han
  et~al.}{2018}]{guofu2018characteristics}
Han, Y., A.~He, D.~Rapach, and G.~Zhou (2018).
\newblock {What Firm Characteristics Drive US Stock Returns?}
\newblock Technical report, Washington University in St. Louis.

\bibitem[\protect\citeauthoryear{Hansen and Jagannathan}{Hansen and
  Jagannathan}{1991}]{HJ1991JPE}
Hansen, L. and R.~Jagannathan (1991).
\newblock {Implications of security market data for models of dynamic
  economies}.
\newblock {\em Journal of Political Economy\/}~{\em 99}, 225--262.

\bibitem[\protect\citeauthoryear{Harvey, Liu, and Zhu}{Harvey
  et~al.}{2016}]{harvey2016and}
Harvey, C.~R., Y.~Liu, and H.~Zhu (2016).
\newblock ... and the cross-section of expected returns.
\newblock {\em Review of Financial Studies\/}~{\em 29\/}(1), 5--68.

\bibitem[\protect\citeauthoryear{He, Feng, Wang, and Wu}{He
  et~al.}{2021}]{he2021predicting}
He, X., G.~Feng, J.~Wang, and C.~Wu (2021).
\newblock {Predicting Individual Corporate Bond Returns}.
\newblock Technical report, City University of Hong Kong.

\bibitem[\protect\citeauthoryear{Heaton, Polson, and Witte}{Heaton
  et~al.}{2017}]{heaton2017deep}
Heaton, J., N.~Polson, and J.~H. Witte (2017).
\newblock Deep learning for finance: deep portfolios.
\newblock {\em Applied Stochastic Models in Business and Industry\/}~{\em
  33\/}(1), 3--12.

\bibitem[\protect\citeauthoryear{Hou, Xue, and Zhang}{Hou
  et~al.}{2020}]{hou2017replicating}
Hou, K., C.~Xue, and L.~Zhang (2020).
\newblock Replicating anomalies.
\newblock {\em Review of Financial Studies\/}~{\em 33\/}(5), 2019--2133.

\bibitem[\protect\citeauthoryear{Jegadeesh and Titman}{Jegadeesh and
  Titman}{1993}]{jegadeesh1993returns}
Jegadeesh, N. and S.~Titman (1993).
\newblock Returns to buying winners and selling losers: {I}mplications for
  stock market efficiency.
\newblock {\em Journal of Finance\/}~{\em 48\/}(1), 65--91.

\bibitem[\protect\citeauthoryear{Jensen, Kelly, and Pedersen}{Jensen
  et~al.}{2022}]{jensen2022there}
Jensen, T.~I., B.~T. Kelly, and L.~H. Pedersen (2022).
\newblock Is there a replication crisis in finance?
\newblock {\em Journal of Finance, Forthcoming\/}.

\bibitem[\protect\citeauthoryear{Jiang, Kelly, and Xiu}{Jiang
  et~al.}{2022}]{jiang2022re}
Jiang, J., B.~T. Kelly, and D.~Xiu (2022).
\newblock (re-) imag (in) ing price trends.
\newblock {\em Journal of Finance, Forthcoming\/}.

\bibitem[\protect\citeauthoryear{Kaniel, Lin, Pelger, and
  Van~Nieuwerburgh}{Kaniel et~al.}{2022}]{kaniel2022machine}
Kaniel, R., Z.~Lin, M.~Pelger, and S.~Van~Nieuwerburgh (2022).
\newblock Machine-learning the skill of mutual fund managers.
\newblock Technical report, National Bureau of Economic Research.

\bibitem[\protect\citeauthoryear{Ke, Kelly, and Xiu}{Ke
  et~al.}{2019}]{ke2019predicting}
Ke, Z.~T., B.~T. Kelly, and D.~Xiu (2019).
\newblock Predicting returns with text data.
\newblock Technical report, National Bureau of Economic Research.

\bibitem[\protect\citeauthoryear{Kelly, Palhares, and Pruitt}{Kelly
  et~al.}{2022}]{kelly2020modeling}
Kelly, B.~T., D.~Palhares, and S.~Pruitt (2022).
\newblock Modeling corporate bond returns.
\newblock {\em Journal of Finance, Forthcoming\/}.

\bibitem[\protect\citeauthoryear{Kelly, Pruitt, and Su}{Kelly
  et~al.}{2019}]{kelly2019characteristics}
Kelly, B.~T., S.~Pruitt, and Y.~Su (2019).
\newblock Characteristics are covariances: {A} unified model of risk and
  return.
\newblock {\em Journal of Financial Economics\/}~{\em 134\/}(3), 501--524.

\bibitem[\protect\citeauthoryear{Kim, Korajczyk, and Neuhierl}{Kim
  et~al.}{2021}]{kim2021arbitrage}
Kim, S., R.~A. Korajczyk, and A.~Neuhierl (2021).
\newblock Arbitrage portfolios.
\newblock {\em Review of Financial Studies\/}~{\em 34\/}(6), 2813--2856.

\bibitem[\protect\citeauthoryear{Kozak, Nagel, and Santosh}{Kozak
  et~al.}{2018}]{kozak2018interpreting}
Kozak, S., S.~Nagel, and S.~Santosh (2018).
\newblock Interpreting factor models.
\newblock {\em Journal of Finance\/}~{\em 73\/}(3), 1183--1223.

\bibitem[\protect\citeauthoryear{Kozak, Nagel, and Santosh}{Kozak
  et~al.}{2020}]{kozak2020shrinking}
Kozak, S., S.~Nagel, and S.~Santosh (2020).
\newblock Shrinking the cross-section.
\newblock {\em Journal of Financial Economics\/}~{\em 135\/}(2), 271--292.

\bibitem[\protect\citeauthoryear{Lettau and Pelger}{Lettau and
  Pelger}{2020a}]{lettau2018estimating}
Lettau, M. and M.~Pelger (2020a).
\newblock Estimating latent asset-pricing factors.
\newblock {\em Journal of Econometrics\/}~{\em 218\/}(1), 1--31.

\bibitem[\protect\citeauthoryear{Lettau and Pelger}{Lettau and
  Pelger}{2020b}]{lettau2020factors}
Lettau, M. and M.~Pelger (2020b).
\newblock Factors that fit the time series and cross-section of stock returns.
\newblock {\em Review of Financial Studies\/}~{\em 33\/}(5), 2274--2325.

\bibitem[\protect\citeauthoryear{Lewellen, Nagel, and Shanken}{Lewellen
  et~al.}{2010}]{lewellen2010skeptical}
Lewellen, J., S.~Nagel, and J.~Shanken (2010).
\newblock A skeptical appraisal of asset pricing tests.
\newblock {\em Journal of Financial Economics\/}~{\em 96\/}(2), 175--194.

\bibitem[\protect\citeauthoryear{Light, Maslov, and Rytchkov}{Light
  et~al.}{2017}]{light2017aggregation}
Light, N., D.~Maslov, and O.~Rytchkov (2017).
\newblock {Aggregation of information about the cross section of stock returns:
  A latent variable approach}.
\newblock {\em Review of Financial Studies\/}~{\em 30\/}(4), 1339--1381.

\bibitem[\protect\citeauthoryear{Merton}{Merton}{1973}]{merton1973intertemporal}
Merton, R.~C. (1973).
\newblock An intertemporal capital asset pricing model.
\newblock {\em Econometrica\/}, 867--887.

\bibitem[\protect\citeauthoryear{Moskowitz and Grinblatt}{Moskowitz and
  Grinblatt}{1999}]{moskowitz1999industries}
Moskowitz, T.~J. and M.~Grinblatt (1999).
\newblock Do industries explain momentum?
\newblock {\em Journal of Finance\/}~{\em 54\/}(4), 1249--1290.

\bibitem[\protect\citeauthoryear{Novy-Marx and Velikov}{Novy-Marx and
  Velikov}{2022}]{novy2022betting}
Novy-Marx, R. and M.~Velikov (2022).
\newblock Betting against betting against beta.
\newblock {\em Journal of Financial Economics\/}~{\em 143\/}(1), 80--106.

\end{thebibliography}
}


\clearpage

\begin{sidewaystable}[htbp]
	\caption{Asset Pricing Performance for Individual Stocks} \label{tab: R2_ind}
	\smallskip
	\noindent {\footnotesize This table provides asset pricing performance results for fitting individual stock returns for both in-sample and out-of-sample analysis. 
    We separate the data into five consecutive folds, use each decade of data as the test sample, and implement a deterministic four-fold cross-validation using the rest four decades of data for training the model.
    The model in bold text is selected by cross-validation.
    Criteria for Total and Predictive $R^2$ (in \%) to individual stocks are provided in Equations (\ref{eqn: totalR2_ind}) and (\ref{eqn: predR2_ind}). We offer results for different layers of augmented deep factor models on CAPM and Fama-French five factors in Panels A and B. In addition, we add results for RP-PCA five factors, IPCA five factors, CAPM, and FF5 for comparison.}
\begin{center}
\footnotesize{\resizebox{0.95\textwidth}{!}{
\begin{tabular}{lcccccccccccccccccccccccccccccccc}
\hline
           &  &  &              &            &  &              &            &  &              &            &  &              &            &  &              &            &  &  &              &            &  &             &             &  &              &            &  &              &            &  &             &             \\
           &  &  & \multicolumn{14}{c}{In-Sample}                                                                                                                        &  &  & \multicolumn{14}{c}{Out-of-Sample}                                                                                                                    \\ \cline{4-17} \cline{20-33} 
           &  &  &              &            &  &              &            &  &              &            &  &              &            &  &              &            &  &  &              &            &  &             &             &  &              &            &  &              &            &  &             &             \\
           &  &  & \multicolumn{2}{c}{72-81} &  & \multicolumn{2}{c}{82-91} &  & \multicolumn{2}{c}{92-01} &  & \multicolumn{2}{c}{02-11} &  & \multicolumn{2}{c}{12-21} &  &  & \multicolumn{2}{c}{72-81} &  & \multicolumn{2}{c}{82-91} &  & \multicolumn{2}{c}{92-01} &  & \multicolumn{2}{c}{02-11} &  & \multicolumn{2}{c}{12-21} \\ \cline{4-5} \cline{7-8} \cline{10-11} \cline{13-14} \cline{16-17} \cline{20-21} \cline{23-24} \cline{26-27} \cline{29-30} \cline{32-33} 
           &  &  & Total        & Pred       &  & Total        & Pred       &  & Total        & Pred       &  & Total        & Pred       &  & Total        & Pred       &  &  & Total        & Pred       &  & Total       & Pred        &  & Total        & Pred       &  & Total        & Pred       &  & Total       & Pred        \\ \hline
           &  &  &              &            &  &              &            &  &              &            &  &              &            &  &              &            &  &  &              &            &  &             &             &  &              &            &  &              &            &  &             &             \\
           &  &  &              &            &  &              &            &  &              &            &  & \multicolumn{12}{c}{Panel A: Deep Factors + CAPM}                                                                         &  &              &            &  &              &            &  &             &             \\ \cline{13-24}
           &  &  &              &            &  &              &            &  &              &            &  &              &            &  &              &            &  &  &              &            &  &             &             &  &              &            &  &              &            &  &             &             \\
DL-CAPM-1L &  &  & 6.90         & 0.16       &  & 8.20         & 0.20       &  & 5.67         & 0.08       &  & 7.53         & 0.54       &  & 7.64         & 0.09       &  &  & 8.58         & 0.26       &  & 3.83        & 0.09        &  & 6.22         & 0.03       &  & 5.60         & 0.46       &  & 3.50        & 0.12        \\
\textbf{DL-CAPM-2L} &  &  & \textbf{7.28}         & \textbf{0.45}       &  & \textbf{8.10}         & \textbf{0.22}       &  & \textbf{5.70}         & \textbf{0.08}       &  & \textbf{6.78}         & \textbf{0.08}       &  & \textbf{8.40}         & \textbf{0.92}       &  &  & \textbf{9.59}         & \textbf{0.80}       &  & \textbf{4.01}        & \textbf{0.17}        &  & \textbf{7.05}         & \textbf{0.06}       &  & \textbf{4.52}         & \textbf{0.10}       &  & \textbf{3.90}        & \textbf{0.05}        \\
DL-CAPM-3L &  &  & 6.69         & 0.18       &  & 8.00         & 0.24       &  & 5.62         & 0.44       &  & 6.94         & 0.25       &  & 7.46         & 0.14       &  &  & 9.61         & 0.30       &  & 3.88        & 0.17        &  & 7.37         & 0.39       &  & 4.99         & 0.26       &  & 2.45        & 0.06        \\
DL-CAPM-4L &  &  & 7.21         & 0.50       &  & 8.14         & 0.39       &  & 6.06         & 0.68       &  & 7.73         & 0.69       &  & 7.50         & 0.45       &  &  & 10.12        & 0.91       &  & 4.08        & 0.35        &  & 7.63         & 0.72       &  & 6.01         & 0.53       &  & 3.76        & 0.19        \\
           &  &  &              &            &  &              &            &  &              &            &  &              &            &  &              &            &  &  &              &            &  &             &             &  &              &            &  &              &            &  &             &             \\
           &  &  &              &            &  &              &            &  &              &            &  & \multicolumn{12}{c}{Panel B: Deep Factors + FF5}                                                                          &  &              &            &  &              &            &  &             &             \\ \cline{13-24}
           &  &  &              &            &  &              &            &  &              &            &  &              &            &  &              &            &  &  &              &            &  &             &             &  &              &            &  &              &            &  &             &             \\
DL-FF5-1L  &  &  & 8.04         & 0.73       &  & 9.02         & 0.43       &  & 7.55         & 0.25       &  & 8.30         & 0.32       &  & 8.41         & 0.16       &  &  & 13.42        & 1.29       &  & 5.23        & 0.39        &  & 9.12         & 0.24       &  & 6.51         & 0.30       &  & 5.08        & 0.17        \\
DL-FF5-2L  &  &  & 7.81         & 0.26       &  & 8.91         & 0.29       &  & 7.63         & 0.44       &  & 8.81         & 0.77       &  & 7.86         & 0.11       &  &  & 13.33        & 0.44       &  & 5.13        & 0.17        &  & 9.38         & 0.51       &  & 6.77         & 0.54       &  & 4.88        & 0.11        \\
\textbf{DL-FF5-3L}  &  &  & \textbf{7.87}         & \textbf{0.28}       &  & \textbf{8.73}         & \textbf{0.21}       &  & \textbf{6.91}         & \textbf{0.18}       &  & \textbf{8.16}         & \textbf{0.12}       &  & \textbf{8.99}         & \textbf{0.48}       &  &  & \textbf{13.44}        & \textbf{0.51}       &  & \textbf{4.99}        & \textbf{0.16}        &  & \textbf{8.13}         & \textbf{0.12}       &  & \textbf{6.30}         & \textbf{0.13}       &  & \textbf{5.51}        & \textbf{0.25}        \\
DL-FF5-4L  &  &  & 7.56         & 0.15       &  & 8.78         & 0.26       &  & 6.99         & 0.22       &  & 7.68         & 0.10       &  & 8.44         & 0.22       &  &  & 12.81        & 0.26       &  & 4.89        & 0.08        &  & 8.29         & 0.22       &  & 5.84         & 0.10       &  & 5.37        & 0.21        \\
           &  &  &              &            &  &              &            &  &              &            &  &              &            &  &              &            &  &  &              &            &  &             &             &  &              &            &  &              &            &  &             &             \\
           &  &  &              &            &  &              &            &  &              &            &  & \multicolumn{12}{c}{Panel C: Other Models}                                                                                &  &              &            &  &              &            &  &             &             \\ \cline{13-24}
           &  &  &              &            &  &              &            &  &              &            &  &              &            &  &              &            &  &  &              &            &  &             &             &  &              &            &  &              &            &  &             &             \\
CAPM       &  &  & 1.02         & 0.06       &  & 1.34         & 0.07       &  & 1.31         & 0.05       &  & 0.96         & 0.08       &  & 1.22         & 0.06       &  &  & 2.02         & 0.14       &  & -0.24       & -0.02       &  & 0.69         & 0.08       &  & 1.59         & 0.06       &  & 0.69        & 0.03        \\
FF5        &  &  & 4.79         & 0.19       &  & 5.75         & 0.17       &  & 5.78         & 0.23       &  & 5.67         & 0.19       &  & 5.73         & 0.23       &  &  & 11.19        & 0.26       &  & 3.73        & 0.15        &  & 4.35         & 0.21       &  & 4.18         & 0.19       &  & 4.05        & 0.15        \\
IPCA5       &  &  & 9.59         & 0.95       &  & 10.65        & 0.94       &  & 9.49         & 0.91       &  & 10.46        & 1.10       &  & 11.03        & 1.34       &  &  & 15.47        & 1.49       &  & 7.50        & 1.55        &  & 11.36        & 1.03       &  & 8.92         & 0.71       &  & 6.68        & -0.03       \\
RP-PCA5     &  &  & 7.40         & 0.60       &  & 8.34         & 0.47       &  & 7.23         & 0.40       &  & 8.17         & 0.75       &  & 8.56         & 0.78       &  &  & 11.91        & 0.94       &  & 4.69        & 0.20        &  & 5.22         & 0.40       &  & 5.22         & 0.60       &  & 4.62        & 0.42        \\ \hline
\end{tabular}
}}
\end{center}
\end{sidewaystable}

\clearpage

\begin{sidewaystable}[htbp]
	\caption{In-Sample Asset Pricing Performance for Portfolios} \label{tab: portR2_ins}
	\smallskip
	\noindent {\footnotesize This table provides asset pricing performance results for fitting portfolio returns for in-sample analysis. 
    The model in bold text is selected by cross-validation.
    Criteria for Total, cross-sectional, and Predictive $R^2$ (in \%) to portfolios are provided in Equations (\ref{eqn: totalR2}), (\ref{eqn: csR2}), and (\ref{eqn: predR2}). We offer results for different layers of augmented deep factor models on CAPM and Fama-French five factors in Panels A and B. In addition, we add results for RP-PCA five factors, IPCA five factors, CAPM, and FF5 for comparison.}
\begin{center}
\footnotesize{\resizebox{0.95\textwidth}{!}{
\begin{tabular}{lccccccccccccc}
\hline
                       &            &            &            &            &  &            &           &           &           &  &             &             &               \\
                       &            &            &            & \multicolumn{8}{c}{In-Sample (1972-2011)}                                       &             &               \\ \cline{5-12}
                       &            &            &            &            &  &            &           &           &           &  &             &             &               \\
                       & \multicolumn{4}{c}{Panel A: Deep Factor + CAPM}   &  & \multicolumn{4}{c}{Panel B: Deep Factor + FF5} &  & \multicolumn{3}{c}{Panel C: Other Models} \\ \cline{2-5} \cline{7-10} \cline{12-14} 
                       &            &            &            &            &  &            &           &           &           &  &             &             &               \\
Portfolios             & DL-CAPM-1L & \textbf{DL-CAPM-2L} & DL-CAPM-3L & DL-CAPM-4L &  & DL-FF5-1L  & DL-FF5-2L & \textbf{DL-FF5-3L} & DL-FF5-4L &  & FF5         & IPCA5        & RP-PCA5        \\ \hline
                       &            &            &            &            &  &            &           &           &           &  &             &             &               \\
                       &            &            &            & \multicolumn{8}{c}{Panel A: Total R2}                                           &             &               \\ \cline{5-12}
$5 \times 5$ Size-BE/ME      & 30.6       & \textbf{32.6}       & 31.3       & 34.5       &  & 74.3       & 74.7      & \textbf{74.1}      & 74.5      &  & 72.3        & 52.8        & 60.5          \\
Industry 49            & 10.0       & \textbf{11.3}       & 11.1       & 11.5       &  & 17.5       & 18.1      & \textbf{16.9}      & 18.8      &  & 14.7        & 7.0         & 14.4          \\
Univariate-Sorted (10$\times$1)      & 26.1       & \textbf{26.7}       & 28.2       & 29.1       &  & 43.1       & 45.0      & \textbf{43.7}      & 44.6      &  & 35.6        & 9.9         & 46.2          \\
Bivariate-Sorted (3$\times$5) & 54.7       & \textbf{53.0}       & 52.8       & 51.7       &  & 69.7       & 70.5      & \textbf{67.8}      & 67.6      &  & 49.2        & 67.1        & 63.6          \\ \hline
                       &            &            &            &            &  &            &           &           &           &  &             &             &               \\
                       &            &            &            & \multicolumn{8}{c}{Panel B: Cross-Sectional R2}                                 &             &               \\ \cline{5-12}
$5 \times 5$ Size-BE/ME      & 9.4        & \textbf{90.4}       & 68.3       & 77.8       &  & 59.3       & 96.3      & \textbf{90.2}      & 97.4      &  & 46.0        & 91.5        & 50.1          \\
Industry 49            & 13.4       & \textbf{38.7}       & 26.0       & 28.4       &  & 41.9       & 67.3      & \textbf{24.4}      & 65.1      &  & 44.8        & -2.4        & 33.0          \\
Univariate-Sorted (10$\times$1)      & 64.6       & \textbf{58.0}       & 80.0       & 71.3       &  & 94.0       & 61.1      & \textbf{89.8}      & 97.1      &  & 93.1        & 94.9        & 94.7          \\
Bivariate-Sorted (3$\times$5) & 83.7       & \textbf{82.0}       & 86.3       & 71.0       &  & 94.0       & 68.7      & \textbf{95.7}      & 75.5      &  & 88.7        & 95.6        & 94.0          \\ \hline
                       &            &            &            &            &  &            &           &           &           &  &             &             &               \\
                       &            &            &            & \multicolumn{8}{c}{Panel C: Predictive R2}                                      &             &               \\ \cline{5-12}
$5 \times 5$ Size-BE/ME      & 0.1        & \textbf{-0.8}       & 0.1        & 0.1        &  & 0.2        & 0.2       & \textbf{0.1}       & 0.2       &  & 0.2         & 0.0         & -0.7          \\
Industry 49            & 0.0        & \textbf{-0.9}       & -0.4       & -0.4       &  & -0.1       & -0.4      & \textbf{-1.1}      & -0.4      &  & -0.2        & -0.9        & -1.6          \\
Univariate-Sorted (10$\times$1)      & -0.1       & \textbf{-1.0}       & 0.0        & 0.0        &  & 0.1        & 0.1       & \textbf{-0.1}      & 0.1       &  & 0.1         & -0.6        & -0.4          \\
Bivariate-Sorted (3$\times$5) & -0.1       & \textbf{-0.4}       & 0.0        & 0.1        &  & 0.1        & 0.1       & \textbf{-0.1}      & -0.1      &  & 0.2         & -0.2        & -0.8          \\ \hline
\end{tabular}
}}
\end{center}
\end{sidewaystable}

\clearpage

\begin{sidewaystable}[htbp]
	\caption{Out-of-Sample Asset Pricing Performance for Portfolios} \label{tab: portR2_oos}
	\smallskip
	\noindent {\footnotesize This table provides asset pricing performance results for fitting portfolio returns for out-of-sample analysis. 
    The model in bold text is selected by cross-validation.
    Criteria for Total, cross-sectional, and Predictive $R^2$ (in \%) to portfolios are provided in Equations (\ref{eqn: totalR2}), (\ref{eqn: csR2}), and (\ref{eqn: predR2}). We offer results for different layers of augmented deep factor models on CAPM and Fama-French five factors in Panels A and B. In addition, we add results for RP-PCA five factors, IPCA five factors, CAPM, and FF5 for comparison.}
\begin{center}
\footnotesize{\resizebox{0.95\textwidth}{!}{
\begin{tabular}{lccccccccccccc}
\hline
                       &            &            &            &            &  &            &           &           &           &  &             &             &               \\
                       &            &            &            & \multicolumn{8}{c}{Out-of-Sample (2012-2021)}                                   &             &               \\ \cline{5-12}
                       &            &            &            &            &  &            &           &           &           &  &             &             &               \\
                       & \multicolumn{4}{c}{Panel A: Deep Factor + CAPM}   &  & \multicolumn{4}{c}{Panel B: Deep Factor + FF5} &  & \multicolumn{3}{c}{Panel C: Other Models} \\ \cline{2-5} \cline{7-10} \cline{12-14} 
                       &            &            &            &            &  &            &           &           &           &  &             &             &               \\
Portfolios             & DL-CAPM-1L & \textbf{DL-CAPM-2L} & DL-CAPM-3L & DL-CAPM-4L &  & DL-FF5-1L  & DL-FF5-2L & \textbf{DL-FF5-3L} & DL-FF5-4L &  & FF5         & IPCA5        & RP-PCA5        \\ \hline
                       &            &            &            &            &  &            &           &           &           &  &             &             &               \\
                       &            &            &            & \multicolumn{8}{c}{Panel A: Total R2}                                           &             &               \\ \cline{5-12}
$5 \times 5$ Size-BE/ME      & 10.7       & \textbf{11.6}       & 13.7       & 27.4       &  & 69.2       & 68.4      & \textbf{69.2}      & 69.2      &  & 68.8        & 55.6        & 46.3          \\
Industry 49            & -2.5       & \textbf{-13.5}      & -9.1       & -8.6       &  & -0.8       & -4.3      & \textbf{-2.6}      & -4.0      &  & 0.0         & -7.3        & -0.6          \\
Univariate-Sorted (10$\times$1)      & 17.9       & \textbf{10.5}       & 15.8       & 18.8       &  & 27.7       & 26.2      & \textbf{27.7}      & 27.8      &  & 23.9        & 2.2         & 31.2          \\
Bivariate-Sorted (3$\times$5) & 24.1       & \textbf{23.9}       & 2.0        & 30.8       &  & 51.4       & 50.2      & \textbf{52.5}      & 52.1      &  & 43.1        & 50.6        & 44.9          \\ \hline
                       &            &            &            &            &  &            &           &           &           &  &             &             &               \\
                       &            &            &            & \multicolumn{8}{c}{Panel B: Cross-Sectional R2}                                 &             &               \\ \cline{5-12}
$5 \times 5$ Size-BE/ME      & 4.8        & \textbf{72.7}       & 50.4       & 59.3       &  & 42.7       & 80.1      & \textbf{72.0}      & 82.0      &  & 32.3        & 73.8        & 35.7          \\
Industry 49            & 6.5        & \textbf{19.1}       & 11.6       & 13.7       &  & 20.3       & 35.0      & \textbf{9.3}       & 32.6      &  & 23.1        & -1.9        & 16.6          \\
Univariate-Sorted (10$\times$1)      & 42.2       & \textbf{37.1}       & 55.3       & 47.4       &  & 71.4       & 39.0      & \textbf{64.8}      & 80.8      &  & 70.1        & 72.2        & 71.1          \\
Bivariate-Sorted (3$\times$5) & 64.9       & \textbf{62.5}       & 67.5       & 49.7       &  & 80.0       & 47.8      & \textbf{81.3}      & 95.9      &  & 70.9        & 81.8        & 81.3          \\ \hline
                       &            &            &            &            &  &            &           &           &           &  &             &             &               \\
                       &            &            &            & \multicolumn{8}{c}{Panel C: Predictive R2}                                      &             &               \\ \cline{5-12}
$5 \times 5$ Size-BE/ME      & 0.3        & \textbf{-1.4}       & 0.4        & 0.2        &  & 0.7        & 0.7       & \textbf{0.9}       & 0.7       &  & 0.7         & -0.2        & 1.7           \\
Industry 49            & 0.1        & \textbf{0.4}        & 0.4        & 0.1        &  & 0.4        & 0.7       & \textbf{0.0}       & 0.1       &  & 0.6         & -0.7        & 1.2           \\
Univariate-Sorted (10$\times$1)      & 0.2        & \textbf{-1.5}       & -0.2       & -0.3       &  & 0.0        & 0.1       & \textbf{-0.3}      & -0.3      &  & -0.1        & -3.6        & 1.6           \\
Bivariate-Sorted (3$\times$5) & 0.5        & \textbf{-0.6}       & -0.2       & -0.4       &  & 0.5        & 0.0       & \textbf{0.3}       & 0.4       &  & 0.3         & -1.1        & 1.1           \\ \hline
\end{tabular}}}
\end{center}
\end{sidewaystable}

\clearpage
\begin{table}[htbp]
	\caption{Investment Performance for Deep Factors} \label{tab: investment}
	\smallskip
	\noindent {\footnotesize This table provides investment performance results for the Sharpe Ratios of the MVE portfolio for both in-sample and out-of-sample analysis. 
    We separate the data into five consecutive folds, use each decade of data as the test sample, and implement a deterministic four-fold cross-validation using the rest four decades of data for training the model.
    The model in bold text is selected by cross-validation.
    We offer results for different layers of augmented deep factor models on CAPM and Fama-French five factors. For comparison, we have added results for CAPM, FF5, IPCA five factors, and RP-PCA five factors. 
 }
\begin{center}
\footnotesize{\resizebox{\textwidth}{!}{
\begin{tabular}{lcccccccccccccc}
\hline
           &  &  &         &         &         &           &           &    &   &        &        &       &       &       \\
           &  &  & \multicolumn{5}{c}{In-Sample}                       &    &   & \multicolumn{5}{c}{Out-of-Sample}       \\ \cline{4-8} \cline{11-15} 
           &  &  &         &         &         &           &           &    &   &        &        &       &       &       \\
           &  &  & 72-81   & 82-91   & 92-01   & 02-11     & 12-21     &    &   & 72-81  & 82-91  & 92-01 & 02-11 & 12-21 \\ \hline
           &  &  &         &         &         &           &           &    &   &        &        &       &       &       \\
           &  &  &         &         &         & \multicolumn{6}{c}{Panel A: Deep Factors + CAPM} &       &       &       \\ \cline{7-12}
           &  &  &         &         &         &           &           &    &   &        &        &       &       &       \\
DL-CAPM-1L &  &  & 1.62 & 1.11 & 2.27 & 3.56   & 1.30   &    &   & 1.20   & 1.67   & 2.86  & 2.97  & 1.31  \\
\textbf{DL-CAPM-2L} &  &  & \textbf{2.86} & \textbf{2.03} & \textbf{3.25} & \textbf{2.08}   & \textbf{4.57}   &    &   & \textbf{4.76}   & \textbf{2.96}   & \textbf{3.46}  & \textbf{1.45}  & \textbf{2.49}  \\
DL-CAPM-3L &  &  & 2.57 & 2.37 & 4.01 & 2.75   & 3.32   &    &   & 1.78   & 4.51   & 4.94  & 2.05  & 0.55  \\
DL-CAPM-4L &  &  & 5.62 & 3.67 & 4.50 & 4.75   & 3.71   &    &   & 7.83   & 8.38   & 4.68  & 4.43  & 2.07  \\
           &  &  &         &         &         &           &           &    &   &        &        &       &       &       \\
           &  &  &         &         &         & \multicolumn{6}{c}{Panel B: Deep Factors + FF5}  &       &       &       \\ \cline{7-12}
           &  &  &         &         &         &           &           &    &   &        &        &       &       &       \\
DL-FF5-1L  &  &  & 4.96 & 3.57 & 3.48 & 2.44   & 2.55   &    &   & 5.72   & 4.23   & 2.99  & 2.31  & 1.02  \\
DL-FF5-2L  &  &  & 4.01 & 3.27 & 3.99 & 4.54   & 4.12   &    &   & 4.18   & 4.17   & 3.92  & 3.96  & 0.77  \\
\textbf{DL-FF5-3L}  &  &  & \textbf{3.30} & \textbf{2.37} & \textbf{3.97} & \textbf{3.24}   & \textbf{6.49}   &    &   & \textbf{3.26}   & \textbf{2.54}   & \textbf{4.02}  & \textbf{2.95}  & \textbf{3.00}  \\
DL-FF5-4L  &  &  & 2.49 & 4.44 & 5.22 & 2.59   & 4.73   &    &   & 1.92   & 6.70   & 5.77  & 1.89  & 1.99  \\
           &  &  &         &         &         &           &           &    &   &        &        &       &       &       \\
           &  &  &         &         &         & \multicolumn{6}{c}{Panel C: Other Models}        &       &       &       \\ \cline{7-12}
           &  &  &         &         &         &           &           &    &   &        &        &       &       &       \\
CAPM       &  &  & 0.60    & 0.46    & 0.46    & 0.55      & 0.34      &    &   & 0.05   & 0.56   & 0.57  & 0.20  & 1.16  \\
FF5        &  &  & 1.25    & 1.00    & 1.14    & 1.12      & 1.18      &    &   & 0.43   & 1.54   & 0.90  & 0.89  & 0.76  \\
IPCA5       &  &  & 6.32    & 5.98    & 6.34    & 7.04      & 8.01      &    &   & 7.20   & 10.81  & 8.11  & 5.00  & 3.67  \\
RP-PCA5     &  &  & 2.38    & 1.80    & 1.85    & 2.80      & 2.48      &    &   & 0.93   & 2.03   & 1.03  & 0.47  & 0.15  \\ \hline
\end{tabular}}}
\end{center}
\end{table}

\clearpage

\begin{table}[htbp]
\caption{Spanning Regressions for Deep Factors} \label{tab: spanreg}
\smallskip
\noindent {\footnotesize This table provides spanning regression results for deep factors generated on the CAPM or Fama-French five-factor benchmarks, corresponding to the selected factor models in Table \ref{tab: R2_ind}. 
Alphas (in \%) and betas on Fama-French five factors are provided with statistical significances to each deep factor.
The analysis uses all 50 years of data, while the deep factors are only generated with the first forty years of data.   
Respectively, $***$ is 1\%, $**$ is 5\%, and $*$ is 10\%.
}
\begin{center}
\footnotesize{
\begin{tabular}{lccccccccc}
\hline
                &  &                  &                 &               &               &               &               &  &       \\
                &  &                  &                 & \multicolumn{3}{c}{Panel A: DL + CAPM}        &               &  &       \\ \cline{5-7}
                &  &                  &                 &               &               &               &               &  &       \\
                &  & $\alpha$ (in \%) & $\beta_{MKTRF}$ & $\beta_{SMB}$ & $\beta_{HML}$ & $\beta_{RMW}$ & $\beta_{CMA}$ &  & $R^2$ \\ \hline
DF\_1           &  & 0.62***          & 0.58***         & 0.79***       & -0.06         & -1.12***      & -0.18         &  & 0.59  \\
                &  & (0.24)           & (0.06)          & (0.08)        & (0.11)        & (0.10)        & (0.17)        &  &       \\
DF\_2           &  & 0.56**           & 0.40***         & 0.64***       & 0.00          & -1.30***      & -0.13         &  & 0.54  \\
                &  & (0.23)           & (0.05)          & (0.08)        & (0.10)        & (0.10)        & (0.16)        &  &       \\
DF\_3           &  & 0.07             & 0.48***         & 0.64***       & 0.10*         & 0.20***       & 0.08          &  & 0.64  \\
                &  & (0.11)           & (0.03)          & (0.04)        & (0.05)        & (0.05)        & (0.08)        &  &       \\
DF\_4           &  & 0.01             & 0.25***         & 0.46***       & -0.44***      & -1.56***      & -0.25         &  & 0.57  \\
                &  & (0.23)           & (0.05)          & (0.08)        & (0.10)        & (0.10)        & (0.16)        &  &       \\
DF\_5           &  & 4.62***          & 0.28***         & 0.03          & 0.50***       & 0.32***       & 0.01          &  & 0.20  \\
                &  & (0.17)           & (0.04)          & (0.06)        & (0.08)        & (0.08)        & (0.12)        &  &       \\
TP              &  & 4.04***          & 0.15***         & -0.07         & 0.32***       & 0.22***       & 0.00          &  & 0.13  \\
                &  & (0.15)           & (0.03)          & (0.05)        & (0.07)        & (0.06)        & (0.10)        &  &       \\
TP (with MktRf) &  & 4.03***          & 0.15***         & -0.07         & 0.32***       & 0.22***       & 0.00          &  & 0.13  \\
                &  & (0.15)           & (0.03)          & (0.05)        & (0.07)        & (0.06)        & (0.10)        &  &       \\
                &  &                  &                 &               &               &               &               &  &       \\
                &  &                  &                 & \multicolumn{3}{c}{Panel B: DL + FF5}         &               &  &       \\ \cline{5-7}
                &  &                  &                 &               &               &               &               &  &       \\
                &  & $\alpha$ (in \%) & $\beta_{MKTRF}$ & $\beta_{SMB}$ & $\beta_{HML}$ & $\beta_{RMW}$ & $\beta_{CMA}$ &  & $R^2$ \\ \hline
DF\_1           &  & 0.05             & 0.54***         & 0.69***       & -0.20*        & -1.38***      & -0.38**       &  & 0.59  \\
                &  & (0.25)           & (0.06)          & (0.08)        & (0.11)        & (0.11)        & (0.18)        &  &       \\
DF\_2           &  & 0.62***          & 0.46***         & 0.61***       & -0.10         & -1.36***      & -0.25         &  & 0.56  \\
                &  & (0.24)           & (0.06)          & (0.08)        & (0.11)        & (0.11)        & (0.17)        &  &       \\
DF\_3           &  & 0.04             & 0.50***         & 0.62***       & -0.08         & -1.07***      & -0.22         &  & 0.53  \\
                &  & (0.23)           & (0.05)          & (0.08)        & (0.10)        & (0.10)        & (0.16)        &  &       \\
DF\_4           &  & 0.00             & 0.09**          & -0.06         & 1.14***       & 0.35***       & -0.26*        &  & 0.40  \\
                &  & (0.20)           & (0.05)          & (0.07)        & (0.09)        & (0.09)        & (0.14)        &  &       \\
DF\_5           &  & 3.50***          & 0.13***         & -0.05         & 0.27***       & 0.90***       & 0.26***       &  & 0.45  \\
                &  & (0.13)           & (0.03)          & (0.04)        & (0.06)        & (0.06)        & (0.09)        &  &       \\
TP              &  & 4.41***          & 0.15***         & 0.07*         & -0.02         & 0.20***       & 0.25***       &  & 0.09  \\
                &  & (0.12)           & (0.03)          & (0.04)        & (0.05)        & (0.05)        & (0.08)        &  &       \\
TP (with FF5) &  & 5.46***          & 0.08**          & 0.06          & -0.02         & 0.16**        & 0.25**        &  & 0.03  \\
                &  & (0.14)           & (0.03)          & (0.05)        & (0.06)        & (0.06)        & (0.10)        &  &       \\ \hline
\end{tabular}
}
\end{center}
\end{table}

\clearpage

\begin{table}[htbp]
	\caption{Linear Exposures for Deep Characteristics \label{tab: exposures_char}}
	\smallskip	
	\noindent \footnotesize This table provides the linear exposures (in \%) to the deep characteristics from 60 raw characteristics. We run the firm-level Fama-Macbeth regression using deep characteristics (augmented 2-layer 5-factor model on CAPM) to 60 ranked raw characteristics. The table is sorted by the exposure values to the fifth deep characteristic (5th column), and the top five important raw characteristics are highlighted. 

\begin{center}
\scriptsize{
\begin{tabular}{lllllll}
\hline
Variable          & Description                              & Char \#1                        & Char \#2                         & Char \#3                         & Char \#4                        & Char \#5                        \\ \hline
sue         & Unexpected Quarterly Earnings            & 3.8***                          & \cellcolor[HTML]{C0C0C0}8.0***   & 3.0***                           & 1.1***                          & \cellcolor[HTML]{C0C0C0}10.2*** \\
mom1m       & Short-Term Reversal (1-1 month)          & \cellcolor[HTML]{C0C0C0}-6.5*** & -4.1***                          & -5.0***                          & -2.6***                         & \cellcolor[HTML]{C0C0C0}-6.7*** \\
bm\_ia      & Industry-adjusted Book to Market         & 1.3***                          & 2.8***                           & 1.8***                           & 0.6***                          & \cellcolor[HTML]{C0C0C0}5.0***  \\
zerotrade   & Number of Zero-Trading Dayes (3 month)   & -2.3***                         & -2.1***                          & -0.2***                          & \cellcolor[HTML]{C0C0C0}-4.7*** & \cellcolor[HTML]{C0C0C0}-4.5*** \\
bm          & Book to Market                           & -2.7***                         & -0.4***                          & -5.4***                          & -2.1***                         & \cellcolor[HTML]{C0C0C0}4.2***  \\
me          & Market Equity                            & -0.2***                         & -2.2***                          & -1.2***                          & 0.2***                          & 4.1***                          \\
rdm         & R\&D to Market Capitalization            & 1.3***                          & 2.3***                           & -2.5***                          & 1.8***                          & 3.1***                          \\
std\_dolvol & Std of Dollar Trading Volume (3 month)   & -1.1***                         & -2.5***                          & -1.2***                          & 0.1**                           & -2.7***                         \\
maxret      & Maximum Daily Return                     & 2.2***                          & 0.7***                           & 0.1***                           & 0.2***                          & -2.7***                         \\
abr         & Abnormal Returns around Earnings Dates   & 3.4***                          & 2.3***                           & 1.2***                           & 0.4***                          & 2.6***                          \\
noa         & Net Operating Assets                     & 0.7***                          & 1.4***                           & \cellcolor[HTML]{C0C0C0}7.9***   & -3.1***                         & 2.2***                          \\
baspread    & Bid-Ask Spread (3m)                      & -1.0***                         & -0.0***                          & -0.9***                          & -1.0***                         & -2.1***                         \\
seas1a      & Seasonality                              & -0.1***                         & 0.4***                           & -2.2***                          & 1.9***                          & 2.0***                          \\
beta        & Beta (3m)                                & -0.1***                         & -2.4***                          & -0.3***                          & -2.2***                         & 1.8***                          \\
roa         & Return on Assets                         & 0.7***                          & -0.8***                          & 0.5***                           & -2.5***                         & 1.6***                          \\
cfp         & Cash Flow to Price ratio                 & -0.7***                         & -3.0***                          & 3.7***                           & -1.1***                         & 1.5***                          \\
pm          & Profit Margin                            & 0.3***                          & -0.4***                          & -0.6***                          & \cellcolor[HTML]{C0C0C0}-5.2*** & 1.4***                          \\
lgr         & Growth in Long-Term Debt                 & -0.3***                         & -0.6***                          & 0.4***                           & 0.3***                          & -1.3***                         \\
cash        & Cash Holdings                            & -4.2***                         & 0.1***                           & \cellcolor[HTML]{C0C0C0}-8.1***  & 2.5***                          & 1.2***                          \\
lev         & Leverage                                 & 1.9***                          & 2.6***                           & 4.3***                           & -2.0***                         & 1.0***                          \\
rd\_sale    & R\&D to Sales                            & 1.9***                          & 4.4***                           & 1.9***                           & 3.3***                          & 1.0***                          \\
sp          & Sales to Price                           & 3.2***                          & 1.6***                           & 0.2***                           & -2.4***                         & 0.9***                          \\
ep          & Earnings to Price                        & -1.4***                         & 0.0                              & 0.4***                           & 0.0                             & 0.9***                          \\
rna         & Return on Net Operating Assets           & 0.4***                          & -1.2***                          & 4.6***                           & -4.1***                         & 0.9***                          \\
agr         & Asset Growth                             & -1.5***                         & -5.2***                          & -0.2***                          & -0.3***                         & -0.9***                         \\
sgr         & Sales Growth                             & 0.7***                          & 0.2***                           & 0.6***                           & -0.3***                         & -0.9***                         \\
chcsho      & Change in Shares Outstanding             & 0.5***                          & -0.4***                          & -1.1***                          & 0.9***                          & -0.8***                         \\
op          & Operating Profitability                  & -0.1***                         & -1.8***                          & 0.5***                           & -1.3***                         & 0.8***                          \\
roe         & Return on Equity                         & 0.4***                          & 0.2***                           & 0.2***                           & -0.2***                         & 0.8***                          \\
herf        & Industry Concentration                   & 1.2***                          & 1.5***                           & 2.3***                           & -0.1***                         & -0.7***                         \\ 
adm         & Advertising Expense-to-market            & 4.3***                          & 2.6***                           & 5.5***                           & -1.3***                         & 0.7***                          \\
rvar\_mean  & Return Variance (3 month)                & \cellcolor[HTML]{C0C0C0}5.1***  & \cellcolor[HTML]{C0C0C0}8.0***   & 5.7***                           & 3.2***                          & -0.7***                         \\
hire        & Employee Growth Rate                     & -1.9***                         & -2.9***                          & -1.7***                          & -1.8***                         & -0.6***                         \\
pscore      & Performance Score                        & -2.9***                         & -4.5***                          & -3.6***                          & -1.4***                         & 0.6***                          \\
mom6m       & Momentum 6m (2-6 month)                  & -1.5***                         & -3.1***                          & 0.5***                           & -1.3***                         & -0.5***                         \\ 
dolvol      & Dollar Trading Volume                    & -1.6***                         & \cellcolor[HTML]{C0C0C0}-8.1***  & -4.3***                          & -1.4***                         & -0.5***                         \\
alm         & Asset Liquidity                          & -0.2***                         & -4.1***                          & 0.1***                           & -3.7***                         & 0.5***                          \\
depr        & Depreciation / PP\&E                     & 0.2***                          & 0.8***                           & 0.1***                           & 0.6***                          & 0.5***                          \\ 
ni          & Net Stock Issues                         & 2.4***                          & 1.3***                           & 0.4***                           & 2.6***                          & -0.4***                         \\
dy          & Dividend to Price                        & -4.9***                         & -3.0***                          & \cellcolor[HTML]{C0C0C0}-13.4*** & 1.5***                          & 0.4***                          \\ 
ato         & Asset Turnover                           & -0.2***                         & -2.5***                          & 0.1***                           & -2.9***                         & -0.4***                         \\
rsup        & Revenue Surprise                         & 1.7***                          & 1.7***                           & 3.1***                           & -0.4***                         & 0.4***                          \\
mom60m      & Long-Term Reversal (13-60 month)         & \cellcolor[HTML]{C0C0C0}-5.5*** & -3.4***                          & \cellcolor[HTML]{C0C0C0}-5.9***  & -0.8***                         & 0.4***                          \\
cashdebt    & Cash Flow to Debt                        & -1.0***                         & -4.4***                          & -1.5***                          & \cellcolor[HTML]{C0C0C0}-4.6*** & -0.4***                         \\
chpm        & Change in Profit Margin                  & 3.0***                          & 3.7***                           & 2.0***                           & 1.1***                          & 0.4***                          \\
gma         & Gross Profitability                      & 0.6***                          & -1.0***                          & 4.4***                           & -4.1***                         & 0.4***                          \\
mom36m      & Momentum 36m (13-36 month)               & \cellcolor[HTML]{C0C0C0}-6.5*** & \cellcolor[HTML]{C0C0C0}-12.2*** & -0.3***                          & -0.8***                         & -0.4***                         \\
pctacc      & Percent Accruals                         & -2.7***                         & -1.1***                          & -1.0***                          & 1.1***                          & -0.3***                         \\
rvar\_ff3   & FF3 Residual Variance (3 month)          & 4.4***                          & 5.8***                           & 0.0                              & \cellcolor[HTML]{C0C0C0}4.1***  & 0.3***                          \\
turn        & Shares Turnover                          & -0.1***                         & -0.5***                          & 0.3***                           & 0.0                             & 0.3***                          \\
cinvest     & Corporate Investment                     & 0.7***                          & 0.1***                           & 1.7***                           & -1.0***                         & 0.2***                          \\
std\_turn   & Std of Share Turnover (3 month)          & -2.1***                         & -2.9***                          & 0.4***                           & -1.8***                         & -0.2***                         \\
nincr       & Number of Earnings Increases             & -0.2***                         & 0.4***                           & -0.0                             & 1.8***                          & -0.1***                         \\
mom12m      & Momentum (2-12 month)                    & 0.2***                          & -2.2***                          & 1.0***                           & -2.6***                         & 0.1***                          \\
rvar\_capm  & CAPM Residual Variance (3 month)         & 0.0                             & -2.4***                          & 3.8***                           & -1.9***                         & -0.1                            \\
ill         & Illiquidity (3m)                         & \cellcolor[HTML]{C0C0C0}10.3*** & \cellcolor[HTML]{C0C0C0}10.4***  & \cellcolor[HTML]{C0C0C0}6.1***   & \cellcolor[HTML]{C0C0C0}8.4***  & 0.1***                          \\
acc         & Working Capital Accruals                 & 0.1***                          & -0.9***                          & 0.1***                           & -1.5***                         & 0.0                             \\
me\_ia      & Industry-Adjusted Market Equity          & 0.0                             & 0.6***                           & 3.8***                           & -1.5***                         & -0.0                            \\
grltnoa     & Growth in Long-Term Net Operating Assets & -0.9***                         & -0.8***                          & -0.8***                          & 0.0                             & 0.0                             \\
chtx        & Change in Tax Expense                    & 0.8***                          & 0.2***                           & 1.3***                           & 0.2***                          & -0.0                            \\ 
\hline
\end{tabular}}
\end{center}
\end{table}

\clearpage

\clearpage

\clearpage
\begin{appendices}
	
\clearpage
\section{Optimization Details} \label{app: optim}
This section illustrates the optimization algorithm for our augmented deep factor model. We use the standard Stochastic Gradient Descent (SGD) approach in deep learning to find the optimal set of parameters. Stochastic Gradient Descent takes an iterative approach to find the optimal set of parameters. The algorithm evaluates the first-order derivative of the objective function in Equation (\ref{eqn: objective}) with respect to all parameters. The first-order derivatives are directly available by carefully applying the backward-chain rule to the augmented deep factor model, executed by \texttt{Tensorflow} library. 
	Specifically, let the superscript $(t)$ denote the $t$-th iteration of the algorithm. Let $\boldsymbol{\Theta}$ 
	denote the set of all parameters to estimate in both neural networks generating latent factors and dynamics of $\boldsymbol{\beta}$, and superscript $(t)$ denote the value of parameters at the $t$-th iteration of the algorithm. We use the RMSProp estimator, illustrated in Algorithm \ref{alg: SGD}.
	
	\begin{algorithm}[ht]
	\caption{RMSPropr algorithm for optimizing parameters}\label{alg: SGD}
	\bigskip
	\small
	\begin{algorithmic}[1]
	    \State Require: learning rate $\eta$, small number for numerical robustness $\delta = 10^{-6}$.
	    \State Initialize all parameters $\boldsymbol{\Theta}^{(0)}$ randomly.
	    \State Initialize accumulated squared-gradients $\mathbf{r}$ at zero.
	    \For{$t$ }
	    \State Randomly draw a mini-batch $\mathcal{B}$ set from the training data, sampling on time horizon.
	    \State Calculate the gradient $\mathbf{g}$, evaluated at the minibatch data.
	    $$
	    \mathbf{g}^{(t)} = \frac{1}{|\mathcal{B}|} \nabla_{\boldsymbol{\Theta}} \sum_{i\in \mathcal{B}} \mathcal{L}^{(t)}_\lambda (\text{data}_i)
	    $$
	    \State Update accumulated squared-gradients ($\odot$ is element-wise product of two vectors)
	    $$
	    \mathbf{r}^{(t+1)} \leftarrow \rho \mathbf{r}^{(t)} + (1 - \rho)\mathbf{g}^{(t)} \odot \mathbf{g}^{(t)}
	    $$
	    \State Update parameters
	    $$
	    \boldsymbol{\Theta}^{(t+1)} \leftarrow  \boldsymbol{\Theta}^{(t)} - \frac{\eta}{\sqrt{\delta + \mathbf{r}}} \odot \mathbf{g}
	    $$
	    \EndFor
	\end{algorithmic}
    \end{algorithm}
    
    Note that the algorithm takes a random subset of the training set per iteration to evaluate the gradient. We let index $\mathcal{B} \subset \left\{1,2,\cdots,T\right\}$, denote the selected training samples. This strategy is called a mini-batch for stable performance and better optimization results. Therefore at each iteration, the loss $\mathcal{L}^{(t)}_\lambda$ only involves the mini-batch data 
    \begin{equation*}
    \mathcal{L}^{(t)}_\lambda(\boldsymbol{\Theta}) \defeq \frac{1}{N|\mathcal{B}|} \sum_{t\in\mathcal{B}} \sum_{i=1}^N\left(r_{i,t} - \widehat{r}_{i,t}\right)^2 + \lambda \sum_{l=0}^{L} \sum_{i\neq j} |A^{[l]}_{i,j}|, 
    \end{equation*}
	where $|\mathcal{B}| < T$, and in practice, we set $|\mathcal{B}|=120$; namely, we use a batch of 120 months for training. This mini-batch setting on the time dimension is reasonable for the asset pricing factor model, which we usually assume with no serial correlation. We set the learning rate $\eta = 0.002$ in our studies.
	Also, we set the number of epochs (roughly the number of times SGD explores the whole training set) to be 300. We observe that the loss function decreases significantly in the first few epochs, and become stable after that. Too many epochs for model training can cause over-fitting. In our study, we consider 300 epochs a reasonable number.

\end{appendices}

\end{document}